\documentclass{article}
\usepackage{multicol,headrule,ams2la}
\usepackage{graphicx}
\usepackage{bm}                       
\usepackage{amssymb}                  
\usepackage{amsbsy} 
\usepackage{caption}
\usepackage{url}
\usepackage{subfigure}
\usepackage{float}
\usepackage{verbatim}
\usepackage{setspace}
\usepackage{tabularx}
\usepackage{color}
\usepackage{soul}
\usepackage{epsfig}
\usepackage{epstopdf}

\graphicspath{{figs/}}      
\graphicspath{{figs/}}   

\def\oddhead{{\rm\hfill\protect\small
Towards V2I Age-aware Fairness Access: A DQN
Based Intelligent Vehicular Node Training and
Test Method
\hfill }}
\def\evenhead{{\rm \hfill \protect\small\it Chinese Journal of
Electronics  \rm\hfill 2022}}

\def\RE{\par\vskip1.5\kh\par\centerline{\bf
References}\par\vskip0.5\kh\par}
\def\REF#1{\par\hangindent\parindent\indent\llap{#1\enspace}\ignorespaces}
\DeclareSymbolFont{lettersA}{U}{txmia}{m}{it}
\DeclareMathSymbol{\piup}{\mathord}{lettersA}{25}
\DeclareMathSymbol{\muup}{\mathord}{lettersA}{22}

\begin{document}
\newtheorem{theorem}{Theorem}
\newtheorem{proposition}{Proposition}
\newtheorem{remark}{Remark}
\newtheorem{proof}{Proof}
\def\footnoterule{\kern 1mm \hrule width 12cm \kern 2mm}
\abovedisplayskip=8.0pt plus 2.0pt minus 1.5pt
\belowdisplayskip=8.0pt plus 2.0pt minus 1.5pt
 \def\thefootnote{}
\TagsOnRight
\def\kh{\baselineskip}
\def\pmb#1{\boldsymbol{#1}}
\thispagestyle{empty}


\vskip 2cm

\begin{center}\huge\bf
Towards V2I Age-aware Fairness Access: A DQN Based Intelligent Vehicular Node Training and Test Method
\end{center}
\footnotetext{\footnotesize Manuscript Received June XX, 20XX; Accepted July XX, 20XX.\ This work was supported in part by the National Natural Science Foundation of China (No.\ 61701197), in part by the open research fund of State Key Laboratory of Integrated Services Networks (No.\ ISN23-11), in part by the National Key Research and Development Program of China (No.2021YFA1000504), in part by the 111 Project (No.\ B12018).}

\vskip 4mm

\begin{center}\large\rm
WU Qiong, SHI Shua, WAN Ziyang, FAN Qiang, FAN Pingyi, and ZHANG Cui 
\\\vskip5mm

\footnotetext{\footnotesize \copyright~2022 Chinese Institute
of Electronics. DOI:10.1049/cje.20XX.0X.0XX}
\end{center}

\begin{center}
{\bf Abstract}
\end{center} 

Vehicles on the road exchange data with base station (BS) frequently through vehicle to infrastructure (V2I) communications to ensure the normal use of vehicular applications, where the IEEE 802.11 distributed coordination function (DCF) is employed to allocate a minimum contention window (MCW) for channel access.\ Each vehicle may change its MCW to achieve more access opportunities at the expense of others, which results in unfair communication performance. Moreover, the key access parameters MCW is the privacy information and each vehicle are not willing to share it with other vehicles. In this uncertain setting, age of information (AoI) is an important communication metric to measure the freshness of data, we design an intelligent vehicular node to learn the dynamic environment and predict the optimal MCW which can make it achieve age fairness.\ In order to allocate the optimal MCW for the vehicular node, we employ a learning algorithm to make a desirable decision by learning from replay history data.\ In particular, the algorithm is proposed by extending the traditional DQN training and testing method.\ Finally, by comparing with other methods, it is proved that the proposed DQN method can significantly improve the age fairness of the intelligent node. 
\begin{center}
{\bf Key words}
\end{center}

vehicles, fair, access, AoI, DQN.

\

\begin{center}{\large\bf I.\ Introduction
}\end{center}

With the development of the economy and quality of life, people have an urgent demand for a more comfortable driving experience$^{[1]}$. Therefore, more and more advanced networks and on-board sensing devices have been introduced to enable a large number of applications to meet the demand of users driving on the road$^{[2]}$.\ Vehicles on the road have to exchange data with the base station (BS) frequently through vehicle to infrastructure (V2I) communications to ensure the normal use of these applications$^{[3,4]}$.\ These data may include image streams, video streams, network instructions, and shared files.\ The vehicles adopt the traditional access mechanism IEEE 802.11 distributed coordination function (DCF) with a fixed minimum contention window (MCW) size to access channel at the media access control (MAC) layer$^{[5,6]}$, but it has a drawback that the spectrum utilization is not efficient due to the dynamic change of vehicle density$^{[7-11]}$. In this case, each vehicle may change its MCW to achieve more access opportunities at the expense of others, which results in unfair communication performance$^{[12]}$. Moreover, the key access parameters MCW is the privacy information and each vehicle are not willing to share it with other vehicles. In this uncertain setting, we design an intelligent vehicular node to learn the dynamic environment and predict the optimal MCW to achieve fair communication performance.

Usually, communication performance is evaluated through wireless communication delay, throughput, and quality of service (QoS)$^{[13-16]}$.\ However, these performance metrics may not be able to reflect the freshness of the transmission data that is important to the vehicular applications.\ As a novel communication metric, the age of information (AoI) has received extensive research attention in recent years$^{[17-19]}$.\ The age of information metric is different from the traditional performance metrics.\ It refers to the time interval between the current time and the generation time of the data to be transmitted.\ As compared to the transmission delay, the AoI can better measure the freshness of the transmitted data$^{[20]}$.

In this paper, we consider the dynamic scenario where the number and MCW of vehicles may change over time.\ In the scenario, when a new vehicle enters the coverage area of the BS, the new vehicle neither knows the MCW of other vehicles nor how other vehicles adjust their own MCW. We designed this newly entered vehicle as an intelligent vehicle node and proposed an extended DQN training and test method to predict the optimal MCW which can ensure its age fairness in the network$^{**}$\footnote{$**$Simulation codes are provided to reproduce the results in this paper: \emph{https://github.com/qiongwu86/Age-Fairness}}. The main contributions are summarized as follows.

1) Considering the dynamic and unknown vehicular networks, we design an intelligent vehicular node to learn the environment and predict the optimal MCW which ensures the age fairness of the intelligent node's transmission data in the network.

2) The age fairness utility of the intelligent vehicular node is defined to measure the relationship between the number of vehicles in the network and the age fairness.

3) We establish a training model by defining the state space, action space, and reward mechanism, and propose an extended DQN training and test method to learn and predict the optimal action at each discrete observation time interval.

4) The performance of the proposed access mechanism is evaluated by simulations under different vehicle characteristics, as compared to other baseline methods.

The rest of this paper is organized as follows.\ Section II reviews the related work.\ Section III briefly describes the system model and average AoI of vehicles in the network.\ Section IV defines the age fairness metric and the DRL framework is set up to formulate age fairness problem.\ Section V presents the extensions DQN algorithm on how to learn the optimal action based on the DRL framework.\ We present some simulation results in Section VI and conclude this paper in Section VII.

\

\begin{center}{\large\bf II.\ Related Work
}\end{center}

In this section, we first review the related work on MAC protocol, then we review the existing work on the AoI.

{\bf 1.\ MAC protocol}

There has been a lot of work to improve the MAC protocol.\ Lv \emph{et al}.$^{[21]}$\ proposed a new function to adjust the contention window (CW) in 802.11 networks to extend DCF.\
Celimuge \emph{et al}.$^{[22]}$ proposed a MAC layer protocol that uses the Q-Learning algorithm to adjust the contention window in order to provide an effective channel access scheme for various network conditions.\
Amin \emph{et al}.$^{[23]}$ proposed a new type of MAC protocol based on IEEE 802.11 DCF, which is called adaptive back-off tuning MAC (ABTMAC).\ It also considers the appropriate transmission attempt rate for both cases where the request to send/clear to send mechanism is not used.\
Zhou \emph{et al}.$^{[24]}$ proposed a practical distributed back-off algorithm called adaptive contention window back-off (ACWB), which is suitable for IEEE 802.11 wireless LAN. It maximizes throughput and fairness based on idle back-off interval statistics.\
Andreas \emph{et al}.$^{[25]}$ proposed a modified version of IEEE 802.11p MAC based on reinforcement learning (RL) to reduce the probability of packet collisions and bandwidth wastage.\ At the same time, the transmission delay is kept within an acceptable level.\
Wu \emph{et al}.$^{[26]}$ proposed a routing scheme based on reinforcement learning, which can improve the contention-efficiency of the IEEE 802.11p MAC layer and achieve low latency.

\vspace{-0.5ex}
{\bf 2.\ Age of information}

\vspace{-0.5ex}
Many works also consider the AoI.\ To solve the problem that the frequent cache of updates of internet of things (IoT) sensors may incur considerable energy costs, Wu \emph{et al}.$^{[27]}$ proposed an online cache update scheme to obtain an effective trade-off between average AoI and energy costs.\ Chen \emph{et al}.$^{[28]}$ investigated the AoI-aware radio resource management problem for an expected long-term performance optimization in a Manhattan grid V2V communication network and proposed a proactive algorithm based on the deep recurrent Q-network. Wu \emph{et al}.$^{[29]}$ considered cellular Internet assisted by UAVs, studied the UAV's AoI minimization problem by designing the UAV's trajectory.\ Wang \emph{et al}.$^{[30]}$ studied the problem of minimizing the weighted sum of the AoI and the total energy consumption of IoT devices.\ To minimize the weighted sum of AoI cost and energy consumption, the author proposed a distributed reinforcement learning method to optimize the sampling strategy.\ Han \emph{et al}.$^{[31]}$ proposed a novel algorithm to solve the optimal blind radio resource scheduling problem in orthogonal frequency division multiplexing access (OFDMA) systems to minimize the long-term average AoI.\ Leng \emph{et al}.$^{[32]}$ considered the user scheduling problem on communication sessions, studied the AoI minimization problem in a network composed of energy harvesting transmitters, and formulated an infinite-state Markov decision problem to optimize AoI.\ Yates \emph{et al}.$^{[33]}$ described the timeliness indicators of AoI and proposed general methods for AoI evaluation and analysis applicable to various sources and systems.\ Kadota \emph{et al}.$^{[34]}$ considered a wireless network with a base station and derived the lower limit of AoI performance that can be achieved for any given network operating under any queuing rule.

As mentioned above, currently there is no work to design a scheme to jointly optimize the MAC protocol and the age fairness, which motivates us to conduct this work.

\

\begin{center}{\large\bf III.\ System Model
}\end{center}

In this section, we will describe the system model in detail.\ It includes scenario model and AoI of the vehicle.

{\bf 1.\ Scenario model}

As illustrated in Fig.1, we consider a vehicular network with a BS deployed at the roadside of a one-way highway.\ The number of vehicles moving in the coverage area of the BS is $N_{v}$ and each vehicle is moving towards the same direction.\ We divide the time domain into discrete time intervals ${n = 1, 2, \dots}$, and the duration of each time interval is $T$.\ At the beginning of each observation time interval, the vehicles enter or leave the network according to the Poisson distribution process with the arrival rate ${\lambda_v}$ and the departure rate ${\mu_v}$, respectively.

\vskip 4mm

\centerline{\includegraphics[scale=0.6]{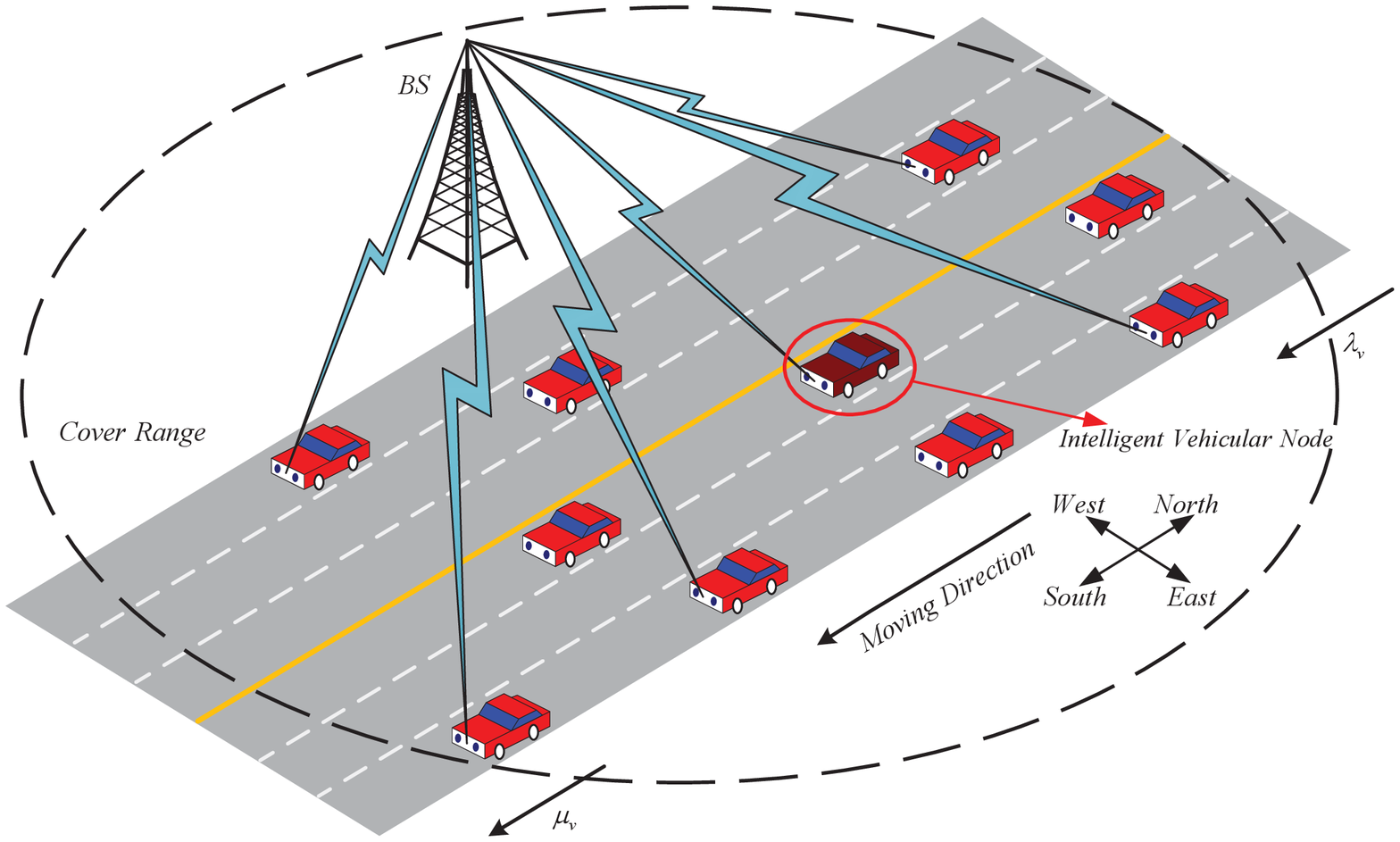} }

\vskip 1mm

\centerline{\footnotesize\begin{tabular}{c} Fig.\ 1.\ System model
\end{tabular}}

\vskip 0.5\baselineskip
Assume that the transceiver is installed on the headstock of each vehicle. Once the headstock of a vehicle enters the coverage of BS, it will transmit the data packet to the BS immediately.\ We consider the network to be operating in a high-load regime$^{[35]}$, \emph{i.e.}, each vehicle always has a packet to transmit, to explore the extreme performance of our scheme.\ Each vehicle employs the IEEE 802.11 DCF mechanism to transmit the packet.\ Specifically, a vehicle initializes a back-off process and randomly chooses an integer value from 0 to $W_0-1$ as the back-off counter, where $W_0$ is the minimum contention window. Then the value of the back-off counter is decreases by 1 after each time slot. When the value of the back-off counter decreases to 0, the vehicle captures the channel. Then the vehicle generates a packet and transmits it to the BS over the captured channel. If more than one vehicle is transmitting at the same time, a collision will occur and thus incurs the transmission failure; otherwise the transmission is successful. The above process will be repeated to transmit different packets.

We assume that the MCWs of all vehicles in the network keep constant in each discrete time interval $n$, \emph{i.e.}, ${\boldsymbol{\omega}^n=[\omega_0^n,\omega_1^n,\dots,\omega_{N_v}^n]}$. The MCWs of these vehicles may be changed at different discrete time intervals.\ We design an intelligent vehicular node, which is referred to as $node \ 0$ in the following of this paper, and adjust its MCW adaptively to achieve the fair age. Note that $node \ 0$ cannot obtain the MCW of any vehicle in the network.\ To achieve the fairness of the AoI, $node \ 0$ communicates with the BS to obtain the observation data, including the number of vehicles, and learns from the replay history data to select the optimal ${\omega_0^n}$ at each time interval.

\vspace{-0.5ex}
{\bf 2.\ Age of information of the vehicle}

\vspace{-0.5ex}
Next, we describe the AoI metric to measure the freshness of data transmitted by a vehicle.\ The process of 802.11 DCF consists of discrete time slots, thus the AoI of vehicle $i$'s data at time slot $t$ is defined as
\begin{equation}
	a_i^t = \max\{t-v_i^t,1\}, \forall i \in \mathcal{N},
	\label{eq1}
\end{equation}
where $v_i^t$ is the time stamp of the generated packet of vehicle $i$, and $\mathcal{N}$ can be expressed as ${\mathcal{N}=\{0,1,\dots,N_v\}}$, where 0 indicates the intelligent vehicular $node \ 0$ and $N_v$ is the number of other vehicles in the network.

To further describe the AoI of vehicle, we assume that at most one packet is cached at each vehicle in the network.\ As shown in Fig.2, a vehicle starts to transmit a packet to the BS at time slot $t_1$, the BS successfully receives the packet transmitted by the vehicle at time slot $t_2$.\ The vehicle will immediately sample and generate the next packet to transmit at the same time.\ Note that the next packet is generated by the vehicle at time slot $t_2$ and arrives at time slot $t_3$ after a time slot.\ At this time, the age of the packet has passed a time slot, so the age at time slot $t_3$ becomes one.\ In addition, other vehicles are transmitting at  time slot $t_4$, thus a collision occurs and the AoI of the vehicle still increases by $1$ after time slot $t_5$.\ The packet is successfully received at the BS at time slot $t_6$, so a new packet will be sampled immediately at the vehicle and the AoI of the vehicle is reset to 1 at time slot $t_7$.\

\vskip 4mm

\centerline{\includegraphics[width=3.3in]{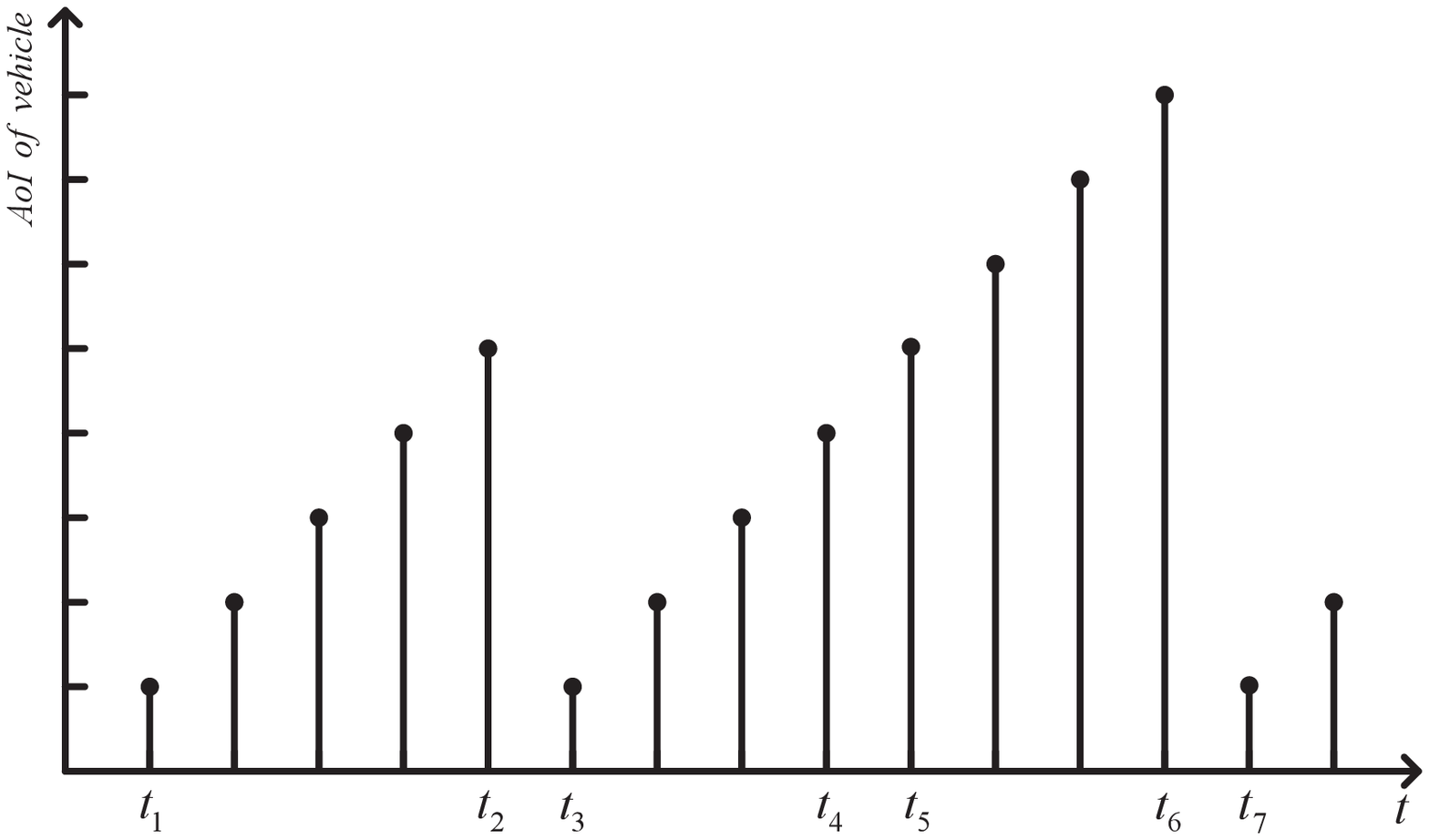} }

\vskip 1mm

\centerline{\footnotesize\begin{tabular}{c} Fig.\ 2.\ Illustration of the AoI of vehicle
\end{tabular}}

\vskip 0.5\baselineskip

Therefore, the AoI of vehicle $i$ at slot $t+1$ can be expressed as:
\begin{equation}
	\resizebox{0.50\hsize}{!}{$
	a_i^{t+1} = 
	\left\{
	\begin{array}{lcl}
		&a_i^{t}+1, &BS \  not \  receives \  a \  packet \  at \  slot \  t, \\
		&1, &BS \ receives \  a \  packet \  at \  slot \  t.
	\end{array}
	\right.
	\label{eq2}$}
\end{equation}

Therefore, we can get the average AoI of vehicle $i$ in an observation time interval as
\begin{equation}
	\overline{\Delta}_i=\frac{T_{slot}}{T}\sum_{t=1}^{T/T_{slot}}a_i^t, \forall i \in \{0,1,\dots,N_v\},
	\label{eq3}
\end{equation}
where $T_{slot}$ is the duration time of a time slot, $T$ is the duration of an observation time interval.

\

\begin{center}{\large\bf IV.\ Problem Formulation
}\end{center}

In order to achieve the age fairness, we use the DRL framework with state, action and reward mechanisms to model the problem of the search for the optimal MCW for $node \ 0$.\ Next, state $s_n$, action $a_n$ and reward $r_n$ for $node \ 0$ at observation time interval $n$ will be defined, respectively.\ The parameters used in this paper are listed in Table.\ 1.

{\bf 1.\ State}

The intelligent $node \ 0$ communicates with the BS within each time interval to obtain the observation information. Since the BS can know the number of vehicles in the network within each time interval $n$, which is denoted as $N_v^n$, through communicating with vehicles, $node \ 0$ can know $N_v^n$ from the BS.\ Moreover, the BS calculates the average AoI of all vehicles in each time interval, thus $node$ 0 can observe the average AoI of all vehicles for each time interval $n$, which is noted as ${\overline{\Delta}_v^n}$. In addition, $node \ 0$ can monitor the channel to observe its average AoI ${\overline{\Delta}_0^n}$ in time interval $n$. At this point, we can define the observation vector ${\boldsymbol{\overline{\Delta}}^n=[\overline{\Delta}_0^n,\overline{\Delta}_v^n]}$ of the AoI in observation interval $n$.\ Since the AoI of a vehicle is affected by the MCW of all vehicles in a time interval and each vehicle may change its MCW at each time interval, the age observation vector $\boldsymbol{\overline{\Delta}}^n$ is a random vector for each time interval.\ Besides, $node \ 0$ can also know its own MCW ${\omega_0^n}$ in time interval $n$.

As described above, the state of $node \ 0$ in the $n$th time interval can be defined as
\begin{equation}
	s_n=[\overline{\Delta}_0^n,\overline{\Delta}_v^n,\omega_0^n,N_v^n],
	\label{eq4}
\end{equation}
where $\overline{\Delta}_0^n$ and $\overline{\Delta}_v^n$ depends on $\omega_0^n$ and the MCW $\omega_i^n$ of all vehicles in the network.\ Since $\overline{\Delta}_0^n$, $\overline{\Delta}_v^n$, $\omega_0^n$ and $N_v^n$ are the values within discrete space, the state space of $node \ 0$ is discrete.

{\tabcolsep=2.5pt \footnotesize
	\begin{center}
		\begin{tabular}{|c|p{10.8cm}|}
			\multicolumn{2}{c}{\bf Table.\ 1.\  Notations used in this paper} \\ \hline
			Notation &\multicolumn{1}{c|}{Description} \\ \hline
			$a_{i}^t$ & The AoI of vehile $i$ at time slot $t$. \\ \hline
			$a_{i}^{t+1}$ & The AoI of vehicle $i$ at time slot $t+1$. \\ \hline
			$a_{n}$ & The action of $node \ 0$ in observation interval $n$. \\ \hline
			$\mathcal{A}$ & The action set. \\ \hline
			$F_{loss}$ &The loss of fairness. \\ \hline
			$m$ &The maximum back-off stage. \\ \hline
			$n$ &The discrete observation interval. \\ \hline
			$N_d$ &The number of vehicular nodes in network. \\ \hline
			$N_{\max}$ &The maximum number of vehicles in network. \\ \hline
			$N_v$ &The number of vehicles in network. \\ \hline
			$Q_{\theta}(s_n,a_n)$ & The state action value of prediction network. \\ \hline
			$Q_{\theta'}(s_n,a_n)$ & The state action value of target network. \\ \hline
			$r_{n}$ & The reward of $node \ 0$ in observation interval $n$. \\ \hline
			$\mathcal{R}$ & The reward set. \\ \hline
			$s_{n}$ & The state of $node \ 0$ in observation interval $n$. \\ \hline
			$\mathcal{S}$ & The state set. \\ \hline
			$T$ & The duration of each observation interval. \\ \hline
			$T_s$ & The average time of a successful transmission. \\ \hline
			$T_c$ & The average time of a failed transmission. \\ \hline
			$T_{slot}$ & The duration time of a time slot. \\ \hline
			$v_i^t$ & The time stamp of the generated packet of vehicle $i$. \\ \hline
			$W_0$ & The contention window of a vehicle. \\ \hline
			$\boldsymbol{\overline{\Delta}}^n$ &\multicolumn{1}{m{10.8cm}|}{The observation vector of average AoI at observation intervals $n$.} \\ \hline
			$\overline{\Delta}_0^n$ &\multicolumn{1}{m{10.8cm}|}{ The average AoI for $node \ 0$ at observation interval $n$.} \\ \hline
			$\overline{\Delta}_v^n$ & \multicolumn{1}{m{10.8cm}|}{The average AoI for all vehicles at observation interval $n$.} \\ \hline
			$\lambda_v$ & The arrival rate of vehicles. \\ \hline
			$\mu_v$ & The departure rate of vehicles. \\ \hline
			$\mu(\overline{\boldsymbol{\Delta}}^n,N_d)$ & The age fairness utility. \\ \hline
			$\boldsymbol{\omega}^n$ &\multicolumn{1}{m{10.8cm}|}{ The MCW vector for all vehicles in the network at observation interval $n$.} \\ \hline
			$\omega_0^n$ &\multicolumn{1}{m{10.8cm}|}{ The MCW of $node \ 0$ at observation interval $n$.} \\ \hline
			$\omega_i^n$ &\multicolumn{1}{m{10.8cm}|}{ The MCW of vehicle $i$ at observation interval $n$.} \\ \hline
			$\alpha$ & The learning rate. \\ \hline
			$\beta$ & The discount factor of future utility. \\ \hline
		\end{tabular}
\end{center}}

{\bf 2.\ Action}


The intelligent $node \ 0$ changes MCW size for the next observation interval $n+1$ according to the $s_n$ locally observed in the observation interval $n$, so the MCW action space of $node \ 0$ can be defined as
\begin{equation}
	a_n = \{2^{j-1}32\}_{j=1}^5 \cup \{2^{j-1}48\}_{j=1}^2.
	\label{eq5}
\end{equation}
In order to make $node \ 0$ more selective, its action space not only covers the limited state space $\Omega$ of the MCW that can be used by the vehicle, but also can extend to additional MCW options: 48 and 96.

{\bf 3.\ Reward}

We first define the age fairness utility for $node \ 0$, then set the reward according to the age fairness utility.

When the network is considered as a saturated state, there are $N_v^n$ vehicles in the network in the time interval $n$.\ If the absolute fairness is provided, theoretically, the proportion of the AoI of nodes in the network should be $1/N_d^n$, where $N_d^n$ is the number of nodes (both $node \ 0$ and other vehicles) in the network, \emph{i.e.}, $N_d^n=N_v^n+1$.\ However, in the real scenario, the proportion of AoI for each node will be given by ${\overline{\Delta}_0^n/(\overline{\Delta}_0^n+\overline{\Delta}_v^n)}$, the absolute difference between ${\overline{\Delta}_0^n/(\overline{\Delta}_0^n+\overline{\Delta}_v^n)}$ and $1/N_d$ can be regarded as the fairness loss.\ According to the relationship between the average age observation vector $\boldsymbol{\overline{\Delta}}^n$ and the number of nodes $N_d^n$ in the network, the fairness loss at time interval $n$ can be defined as
\begin{equation}
	F_{loss}^n = \left|\frac{\overline{\Delta}_0^n}{\overline{\Delta}_0^n+\overline{\Delta}_v^n}-\frac{1}{N_d^n}\right|.
	\label{eq6}
\end{equation}

According to Eq.(6), the age fairness utility at time interval $n$ can be defined as
\begin{equation}
	\begin{aligned}
		\mu(\overline{\boldsymbol{\Delta}}^n,N_d^n) &= 1-F_{loss}^n\\
		&= 1-\left|\frac{\overline{\Delta}_0^n}{\overline{\Delta}_0^n+\overline{\Delta}_v^n}-\frac{1}{N_d^n}\right|.
		\label{eq7}
	\end{aligned}
\end{equation}

In this paper, as the number of vehicles and MCW changes, $node \ 0$ aims to maintain its age fairness utility in the network, so the reward of $node \ 0$ at the observation interval $n$ can be defined as its age fairness utility, i.e.,
\begin{equation}
	r_n = \mu(\overline{\boldsymbol{\Delta}}^n,N_d^n).
	\label{eq8}
\end{equation}

With the above definition, we will further introduce the MCW optimization problem for $node \ 0$ in the vehicle environment.\ Specifically, given the age observation vector for the observation interval $n$, \emph{i.e.}, ${\boldsymbol{\overline{\Delta}}^n=[\overline{\Delta}_0^n,\overline{\Delta}_v^n]}$, the MCW of $node \ 0$ for observation interval n, \emph{i.e.}, $\omega_0^n$, and the number of nodes $N_d^n$, $node \ 0$ predicts and allocates its MCW for the next observation interval $n+1$, \emph{i.e.}, $\omega_0^n+1$, to maximize the long-term reward, \emph{i.e.}, the age fairness utility of the entire observation process $ \mathbb{E}\left[\sum_{k=n}^{\infty}\beta^{k-n}\mu(\overline{\boldsymbol{\Delta}}^k,N_d^n)\right]$, where $\beta \in (0, 1)$ is used to establish the importance of future utility.

{\bf 4.\ Policy}

The action taken by $node \ 0$ in each state $a_n = \pi(s_n)$ is determined by the policy $\pi$ that maps from the state set $\mathcal{S}$ to the action set $\mathcal{A}$.\ The goal of Q-learning is to find the optimal policy $\pi^*$ to maximize the long-term expected cumulative discount reward.\ To do this, we define the optimal Q function $Q^*: \mathcal{S} \times \mathcal{A} \to \mathcal{R}$, where $Q^*(s_n, a_n)$ corresponds to the long-term reward when choosing action $a_n$ in state $s_n$, \emph{i.e.}, following the optimal policy $\pi^*$.\ Therefore, the optimal policy can be defined as
\begin{equation}
	\pi^*(s_n)=\mathop{arg\!\max}\limits_{a_n} \, Q^*(s_n,a_n).
	\label{eq9}
\end{equation}

\

\begin{center}{\large\bf V.\ Solution
}\end{center}

Since the state and action space are discrete, the DQN algorithm is more suitable to solve the DRL problems.\ Therefore, we propose an extended DQN algorithm to obtain the optimal action policy. In this section, we first introduce how we extend the DQN training and testing method to better solve the above problem, then describe the training procedures to obtain the optimal age fairness utility, finally introduce how to test the performance under the optimal policy.

{\bf 1.\ Extensions to DQN}

We extend the traditional DQN from five aspects, i.e., double Q-learning, dueling networks, multi-step learning, distributional RL and noisy nets. The flow diagram of the extended DQN algorithm is shown in Fig.3.

{\bf 1).\ Double Q-learning}

The classical method of searching for the optimal Q value is based on the value iteration method of Bellman equation$^{[36]}$, \emph{i.e.},
\begin{equation}
	\begin{aligned}
		& Q_{\theta}(s_n,a_n) \leftarrow Q_{\theta}(s_n,a_n) + \\
		& \alpha\left(r_{n}+\gamma\mathop{\max}\limits_{a_{n+1}}\,Q_{\theta}(s_{n+1},a_{n+1})-Q_{\theta}(s_n,a_n)\right),
		\label{eq10}
	\end{aligned}
\end{equation}
where $Q_{\theta}(s_n,a_n)$ is the state action value when action $a_n$ is selected under state $s_n$, $\alpha$ is the learning rate, $r_{n}$ is the immediate reward obtained by taking action $a_{n}$ under state $s_{n}$, $\gamma$ is the discount factor, $Q_{\theta}(s_n+1,a_n+1)$ is the sta-

\vskip 4mm

\centerline{\includegraphics[scale=0.9]{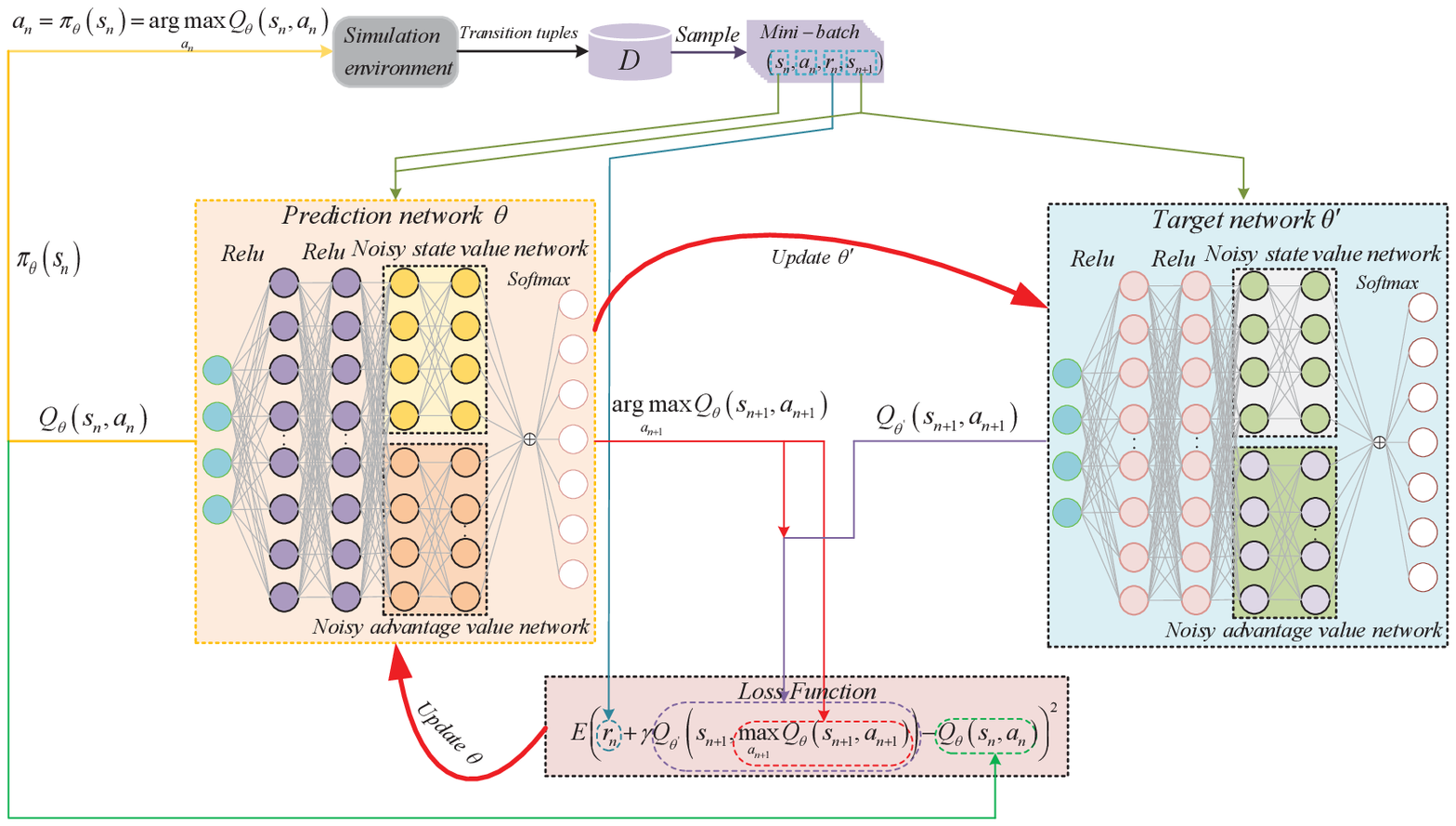} }

\vskip 1mm

\centerline{\footnotesize\begin{tabular}{c} Fig.\ 3.\ Flow Diagram of Extended DQN
\end{tabular}}

\vskip 0.5\baselineskip

\noindent te action value when action $a_{n+1}$ is selected under state $s_{n+1}$.\ However, only one neural network parameter $\theta$ is used in Eq.(10) for Q value iteration, which results in the overestimation bias, thus it affects the Q value and further affects the performance of learning.\ Double Q-learning is designed in calculating the target Q value with two different neural network parameters including the prediction network parameter $\theta$ and the target network parameter $\theta'$$^{[37]}$.\ The particular formula is shown as follows, 
\begin{equation}
	\resizebox{0.50\hsize}{!}{$
	\begin{aligned}
		& Q_{\theta}(s_n,a_n) \leftarrow Q_{\theta}(s_n,a_n) + \\
		& \alpha\left(r_{n}+\gamma Q_{\theta'}(s_{n+1},\mathop{\max}\limits_{a_{n+1}}\,Q_\theta(s_{n+1},a_{n+1}))-Q_{\theta}(s_n,a_n)\right).
		\label{eq11}
	\end{aligned}$}
\end{equation}

The prediction network is responsible to select the action and the target network is used to calculate the target Q value.\ By decoupling the selection of actions and the estimation of the value of state actions, the harmful overestimation of DQN is reduced, thereby improving the stability of the algorithm$^{[38]}$.

{\bf 2).\ Dueling networks}

Different from the traditional DQN where the neural network directly outputs the Q value of each action, the dueling DQN first divides the fully connected layer of the network into an output state value $V(s_n)$ and an output action advantage value $A(s_n, a_n)$.\ The state value function $V(s_n)$ has nothing to do with actions, but represents the value of the static state environment itself.\ Meanwhile, action advantage function $A(s_n,a_n)$ is related to the action and represents the additional value of choosing an action.\ In addition, the action advantage value function can reflect the difference between the value obtained by taking an action and the average value obtained by this state.\ Finally, the two features are aggregated in a fully-linked manner and nonlinear activation is performed through the $Softmax$ activation function to obtain the Q value of each action.\ In practical applications, the action advantage function is generally set as the single action advantage function minus the average value of all the action advantage functions in a certain state, \emph{i.e.},
\begin{equation}
	\begin{aligned}
		& Q_{\theta}(s_n,a_n,\alpha^n,\beta^n)= \\
		& V_\theta(s_n,\beta^n) +A_\theta(s_n,a_n,\alpha^n)-\frac{\sum\limits_{a_n'}{A_\theta\left(s_n,a_n^{'},\alpha^n\right) }}{N_{actions}},
		\label{eq12}
	\end{aligned}
\end{equation}
where $\alpha^n$ and $\beta^n$ represent the number of two fully linked layers of the neural network, respectively, $V_\theta(s_n,\beta^n)$ is the state value of state $s_n$, $A_\theta(s_n,a_n,\alpha^n)$ is the action advantage value when action $a_n$ is selected under state $s_n$, $N_{actions}$ is the number of actions provided by the neural network, $a_n'$ is all available actions.\ Based on this, $node \ 0$ can finally achieve a more accurate $V(s_n,\beta^n)$ in the environment without the influence of actions, \emph{i.e.}, it can narrow the range of Q value, remove redundant degrees of freedom and further improve the stability of the algorithm$^{[39]}$.

{\bf 3).\ Multi-step learning}

The traditional DQN uses the immediate reward $r_n$ and the value estimate at the next step as the target value.\ However, with a large number of deviations in the network parameters at the early stage, the target value obtained by this method will also require a large number of deviations, and thus incur a relatively slow learning rate$^{[40]}$.\ A more accurate estimate can be obtained by changing single-step learning to multi-step learning, because immediate rewards can be accurately obtained through interaction with the environment, so the immediate reward can be rewritten as
\begin{equation}
	r_{n}^{(k)}\equiv\sum_{j=0}^{k-1}\gamma_r \ r_{n+j},
	\label{eq13}
\end{equation}
where $\gamma_r$ is the discount factor of immediate reward.\ We can replace $r_n$ in Eq.(11) with  $r_{n}^{(k)}$ Eq.(13) to minimize the loss of action value.\ Note that, proper adjustment of the number of steps in multi-step learning can achieve faster learning.

{\bf 4).\ Distributional RL}

In the traditional DQN, the outputs of the network are always the expected estimated value of the state-action value Q.\ However, the expectations for different groups of state-action values may be the same, we should choose a more stable action if we want to reduce the risk of taking actions.\ Distributed DQN based deep reinforcement learning models are designed from a distributed perspective.\ It uses a histogram to represent the estimate of the value distribution, limits the value among $[v_{min}, v_{max}]$ and selects $N$ equidistant value sampling points in $[v_{min},v_{max}]$.\ Thus the output of the network is the probability of these $N$ value sampling points, where the sampling point with the largest probability is the optimal action to be selected.\ After projecting the state-action value output by the neural network onto the vector $\boldsymbol{z}$, the action with the highest probability becomes the stable action we expect.\ Among them, the calculation method of each element in $\boldsymbol{z}$ is
\begin{equation}
	z^i=v_{\min}+(i-1)\frac{v_{\max}-v_{\min}}{N-1}, i \in \{1, \dots, N\}.
	\label{eq14}
\end{equation}

{\bf 5).\ Noisy nets}

In the learning process, $node \ 0$ needs to perform a lot of action explorations, but the $\varepsilon-greedy$ algorithm has limitations in performing exploration$^{[41]}$.\ In order to improve the search ability of the agent, by combining the determinism and the noise linear hidden layer of the noise flow, \emph{i.e.}, the weight and the bias are interfered by some parameter zero-mean noise, the existing expression $y=wx+b$ of the neural network hidden layer can be modified as
\begin{equation}
	y=(\mu^w+\sigma^w \odot \varepsilon^w)x+(\mu^b+\sigma^b \odot \varepsilon^b).
	\label{eq15}
\end{equation}

Here, $\odot$ denotes the element-wise product, $\mu^w$ and $\mu^b$ obey a uniform distribution ranging from ${[-\frac{1}{\sqrt{N_n}},\frac{1}{\sqrt{N_n}}]}$, $N_n$ is the number of neural network nodes, $\sigma^w$ and $\sigma^b$ can be initialized via $\frac{S_{td}}{\sqrt{N_n}}$, and $S_{td}$ is the neural network parameter.\ In addition, $\varepsilon^b$ and $\varepsilon^w$ are the constant randomly generated noise variables in each exploration, which can be expressed as
\begin{equation}
	\left\{
	\begin{array}{lcl}
		&\varepsilon^w=f(\varepsilon_i), \\
		&\varepsilon^b=f(\varepsilon_j),
	\end{array}
	\right.
	\label{eq16}
\end{equation}
where $\varepsilon_i$ and $\varepsilon_j$ follow the standard normal distribution with mean 0 and variance 1, $f(x)$ can be expressed as ${f(x)=sgn(x)\sqrt{|x|}}$.\ Over time, the network can overcome different rates of noise flow through self-learning, thereby completing the exploration of actions in the form of self-annealing.

{\bf 2.\ Training stage}

The pseudocode of the proposed algorithm is described in Algorithm 1.\ For ease of understanding, we will further introduce the proposed DQN algorithm in detail below in conjunction with the pseudocode.

\vspace{-0.5em}
{\tabcolsep=4pt\small
	\begin{center}
		\begin{tabular}{lp{82mm}} \hline
			\multicolumn{2}{l}{{\bf Algorithm 1}~~ Training Stage for Extended DQN}\\ \hline
			&\hspace*{-3mm}1: Create simulation environment, initialize $N_v$ according to the $\lambda_v$ and $\mu_v$, initialize $\omega_i^n$;\\
			&\hspace*{-3mm}2: Create predict network and target network;\\
			&\hspace*{-3mm}3: Initialize replay experience buffer $\mathcal{D}$ to capacity $\left| \mathcal{D} \right|$;\\
			&\hspace*{-3mm}4: Reset simulation environment;\\
			&\hspace*{-3mm}5: FOR episode = $1$ to $E_{max}$\\
			&\hspace*{-3mm}6: \qquad Initialize observation sequence
			$s_0=[\overline{\Delta}_0^0,\overline{\Delta}_v^0,\omega_0^0,N_v]$\\
			&\hspace*{0.5mm}  \qquad and normalized sequenced $\phi_0 = \phi(s_0)$;\\
			&\hspace*{-3mm}7: \qquad FOR time interval $n$ = $1$ to $T_{max}$\\
			&\hspace*{-3mm}8: \hspace*{0.5mm}\qquad\qquad Generate the action $a_n$ according to Eq.(9);\\
			&\hspace*{-3mm}9: \hspace*{0.5mm}\qquad\qquad Execute next state $s_{n+1}$ and observe reward $r_{n}$\\
			&\hspace*{0.5mm}   \qquad\qquad from the system model;\\
			&\hspace*{-3mm}10:\hspace*{-1mm} \qquad\qquad Normalized sequenced $\phi_{n+1} = \phi(s_{n+1})$;\\
			&\hspace*{-3mm}11:\hspace*{-1mm} \qquad\qquad Store transition $(\phi_{n},a_{n},r_{n},\phi_{n+1})$ in $\mathcal{D}$;\\
			&\hspace*{-3mm}12:\hspace*{-1mm} \qquad\qquad IF number of tuples in $\mathcal{D} \geq I$ THEN\\
			&\hspace*{-3mm}13: \qquad\qquad\quad Randomly sample a mini-batch of $I$ transi-\\
			&\hspace*{2mm}   \qquad\qquad\quad tions tuples from $\mathcal{D}$;\\
			&\hspace*{-3mm}14: \qquad\qquad\quad Update the predict network by minimizing\\
			&\qquad\qquad\qquad the loss function according to Eq.(17);\\
			&\hspace*{-3mm}15:\hspace*{-1mm} \qquad\qquad IF time interval $n$ = $T_{max}$ THEN\\
			&\hspace*{-3mm}16: \qquad\qquad\quad Update target networks $\theta' \leftarrow \theta$.\\
			\hline
		\end{tabular}
\end{center}}
First, we create the simulation environment.\ In this step, we declare the available set of MCWs of both $node \ 0$ and vehicles, initialize the MCW value of $node \ 0$ and vehicles at the beginning and initialize the number  of vehicles in the network according to the vehicle arrival rate $\lambda_v$ and departure rate $\mu_v$ in the network.

Then we create neural networks for $node \ 0$.\ We use two neural networks to create our extended DQN network, \emph{i.e.}, the prediction network and the target network.\ The prediction neural network consists of four layers.\ The first two layers are ordinary fully connected layers and the latter two layers are Noise Nets, where each layer is composed of $N_n$ neural network nodes.\ Each layer of the neural network is fully connected and the output of each node uses $ReLu$ activation function for non-linear activation.\ In addition, the input of the neural network is the state, while the output is the state action value of all available actions.\ We also introduce the dueling network to build a neural network. The prediction network starts from the third layer and is divided into a state value network and an action advantage value network.\ The state value network and the action advantage value network share the first two layers of fully connected neural networks.\ The two-way features are aggregated together in a fully-linked manner and the $Softmax$ activation function is used for nonlinear activation before outputting the Q value.\ Then according to our description of distributed RL, the state action value is limited to $[v_{min}, v_{max}]$.\ By selecting $N$ equidistant value sampling points in $[v_{min}, v_{max}]$, the network outputs the projection vector $\boldsymbol{z}$ of these $N$ value sampling points and finally outputs the action index value with the largest probability value, \emph{i.e.}, the most stable action.\ The target network has the same structure as the prediction network and will not be described in detail here.

Then we initialize the experience buffer $\mathcal{D}$ to store the state, action, reward and next state set during the training process.\ The experience buffer has the storage capacity of $\left| \mathcal{D} \right|$ sets, which corresponds to line 3 of the pseudocode.

Before starting the training, we need to reset the environment, which corresponds to line 4 of the pseudocode.

At the beginning of the loop, we initialize the state set for $node \ 0$ and execute it at the beginning of each episode.\ In addition, to facilitate the training process of neural networks and reduce the magnitude of data processing, we have carried out data normalization processing on the state set of $node \ 0$. This corresponds to line 6 of the pseudocode.

For each observation interval $n$, we input the state into the neural network above to decide the action.\ At the initial stage of training, due to the influence of the randomly initialized parameters, the action will be randomly given by the neural network. This corresponds to line 8 of the pseudocode.

In the training process, the MCW of vehicles in the environment will change according to a Markov process and the number of vehicles will change momentarily in each time interval $n$.\ After $node \ 0$ selects the MCW, it obtains the next state $s_{n+1}$ by interacting with the environment and can calculate the immediate reward for each state transition according to Eq.(8).\ This corresponds to line 9 of the pseudocode.

After getting the next state $s_{n+1}$, we also need to execute normalization for the convenience of processing.\ Then we cache the current state $\phi_{n}$, the action $a_{n}$, the next state $\phi_{n+1}$ and the immediate reward $r_{n}$ into the experience buffer $\mathcal{D}$ for subsequent parameter training.

If the number of samples in the experience buffer is less than the mini-batch size, the interaction with the environment will continue and the sample will be cached into the experience buffer.\ On the contrary, a sample is randomly picked up from the experience buffer to train neural network parameters according to the defined loss function$^{[42]}$.\ The loss function is defined as
	\begin{equation}
		\resizebox{0.50\hsize}{!}{$
		\begin{aligned}
			& \mathcal{L}(\theta)= \\
			& \mathbb{E}\left(r_{n}+\gamma Q_{\theta'}(s_{n+1},\mathop{\max}\limits_{a_{n+1}}\,Q_\theta(s_{n+1},a_{n+1}))-Q_\theta(s_n,a_n)\right)^2.
			\label{eq17}
		\end{aligned}$}
	\end{equation}

The neural network parameters are gradually updated through gradient descent by backpropagation, \emph{i.e.}, ${\theta \leftarrow \theta-\alpha \nabla \theta}$, where $\alpha=0.0001$ is the learning rate of gradient descent.\ This corresponds to lines 13 to 14 of the pseudocode.

Finally, at the end of each episode, we will update the parameters $\theta'$ of the target network through the neural network parameters $\theta$ of the prediction network, \emph{i.e.}, ${\theta' \leftarrow \theta}$.\ This corresponds to the end of the pseudocode.

Through continuous iterative calculations, the neural network parameters of double DQN will finally approach the optimal $\theta^*$, and the training process is completed at this time.

{\bf 3.\ Testing stage}

To evaluate the performance of the proposed algorithm, we load the optimal neural network parameters $\theta^*$ obtained in the training stage and start to test its performance.\ The pseudocode of the testing stage is shown in Algorithm 2.

\vspace{-0.5em}
{\tabcolsep=4pt\small
	\begin{center}
		\begin{tabular}{lp{82mm}} \hline
			\multicolumn{2}{l}{{\bf Algorithm 2}~~ Testing Stage for Extended DQN}\\ \hline
			&\hspace*{-3mm}1: Create simulation environment;\\
			&\hspace*{-3mm}2: Import the trained neural network parameters;\\
			&\hspace*{-3mm}3: FOR episode = $1$ to $E_{max}$\\
			&\hspace*{-3mm}4: \qquad Initialize $N_v$ according to the $\lambda_v$ and $\mu_v$;\\
			&\hspace*{-3mm}5: \qquad Initialize $\omega_i^n$;\\
			&\hspace*{-3mm}6: \qquad Initialize initial observation state\\ &\hspace*{1mm}\qquad$s_0=[\overline{\Delta}_0^0,\overline{\Delta}_v^0,\omega_0^0,N_v]$ and normalized sequenced\\
			&\hspace*{1mm}\qquad$\phi_0 = \phi(s_0)$;\\
			&\hspace*{-3mm}7: \qquad FOR time interval $n$ = $1$ to $T_{max}$\\
			&\hspace*{-3mm}8: \qquad\qquad Generate the action according to\\ &\hspace*{1mm}\qquad\qquad$\pi^*(s_n)=\mathop{arg\!\max}\limits_{a_n} \, Q_{\theta^*}^*(s_n,a_n)$;\\
			&\hspace*{-3mm}9: \qquad\qquad Execute action $a_{n}$, observe reward $r_{n}$ and new\\
			&\hspace*{1mm}\qquad\qquad state $s_{n+1}$ from the system model;\\
			\hline
		\end{tabular}
\end{center}}

\

\begin{center}{\large\bf VI.\ Numerical Simulation
}\end{center}

In this section, we evaluate the performance of the algorithm by extensive simulation experiments and discuss the results in detail.\ The simulation parameters are shown in Table. 2.\ The simulation is based on python $3.8$.\ The system scenario is a one-way highway network scenario as described in section III. We have considered two scenarios with dynamic MCW and number of vehicles, \emph{i.e.}, the simple vehicle scenario and the complex vehicle scenario mentioned in the below simulation analysis.\ In these two scenarios, the simulation analysis is carried out when the MCW transition probability $ps$ of all vehicles is $1.0$ and $0.75$ respectively.

\vspace{-0.4cm}
\textbf{{\tabcolsep=2.5pt \footnotesize
		\begin{center}
			\begin{tabular}{|c|c|c|c|}
				\multicolumn{4}{c}{\bf Table.\ 2.\ Related parameters} \\ \hline
				\multicolumn{4}{|c|}{Parameters of System Model} \\ \hline
				Parameter & Value & Parameter & Value \\ \hline
				$N_{v}$ & $0-9$ & $N_d$ & $1-10$ \\ \hline
				$\lambda_v$ & $1-6$ & $\mu_v$ & $1-6$ \\ \hline
				$T$ & $1 s$ & $T_{slot}$ & $50\mu s$ \\ \hline
				$T_s$ & $179.64\mu s$ & $T_{c}$ & $174.26\mu s$ \\ \hline
				$m$ & $3$ & $N_{max}$ & $0-9$ \\ \hline
				\hline
				\multicolumn{4}{|c|}{Parameters of Extended DQN} \\ \hline
				Parameter & Value & Parameter & Value \\ \hline
				$E_{max}$ &$200/1000$ &$T_{max}$ &$200/400$\\ \hline
				$N_n$ &$64/480$ &$S_{td}$ &$0.4$\\ \hline
				$\alpha^n$ &$64/480$ &$\beta^n$ &$64/480$\\ \hline
				$v_{min}$ &$45/43$ &$v_{max}$ &$50$\\ \hline
				$\gamma$ &$0.99$ &$\alpha$ &$0.0001$\\ \hline
				$\gamma_r$ &$0.99$ &$\beta$ &$0.99$\\ \hline
				$\left|\mathcal{D}\right|$ &$10000/100000$ &$I$ &$32$\\ \hline
			\end{tabular}
\end{center}}}

In the simple vehicle scenario, the MCWs of all vehicles follow a Markov process with only two states $[s_1,s_2]$.\ In our simulation experiment, these two states are $[32,128]$, and we set $\omega_i^n \in [32,128], \forall i \neq 0$.\ In the second complex vehicle scenario, the MCWs of all vehicles follow a Markov process with five states $[s_1, s_2, s_3, s_4, s_5]$.\ In our simulation experiment, these five states are $[32,64,128,256,512]$, and we set $\omega_i^n \in [32,64,128,256,512], \forall i \neq 0$.\ At each discrete time interval, the MCWs of all vehicles in the network transition to the next state with probability $ps$, and the state changes follow an increasing order from $s_1$ to $s_5$, and then a decreasing order from $s_5$ to $s_1$, so on and so forth.

{\bf 1.\ Training stage}

We first obtain the age dataset based on real-time protocol simulation.\ We will train the model for $node \ 0$ based on this dataset.\ To make our results more general, we train the model 10 times in different scenarios and averaged the convergence curves.

Fig.4 shows the learning curve of the training process in the simple vehicle scenario, which reflects the average age fairness utility in each step, \emph{i.e.}, discrete time interval, under different episodes.\ It can be seen that the average age fairness utility gradually increases from episode $0$ to episode $60$ when $ps=1.0$.\ Then the age fairness utility turns to be stable from episode $60$ to $200$, which means that the optimal MCW adjustment policy for this simple vehicle scenario has been learned.\ For the convergent curve, we can see that it is not smooth, this is mainly because in the training process of the model, we use the real simulation value of the real-time protocol to calculate the age fairness utility.\ In addition, due to the randomness of the integer selection of the back-off counter, and the impact of back-off freezing, collision transmission, and successful transmission during device communication, there will be a slight difference in the average age statistics for each time interval.\ In addition, for the scenario where the state transition probability $ps=0.75$, it can be seen that the curve has the same convergence trend as the scenario of $ps=1.0$, but the age fairness utility after convergence is lower than that when $ps=1.0$, and the amplitude of its fluctuation is higher than that in $ps=1.0$ scenario.\ This is mainly because when the transition probability is not $1.0$, the changes of MCW states of other vehicles will become complicated and no longer have a regularity to follow.\ At the same time, the number of vehicles changes dynamically at each time interval, which makes the MCW prediction and adjustment of $node \ 0$ challenging.\ Therefore, the prediction may be undesirable in some steps, resulting in a decrease in the average age fairness utility.
\centerline{\includegraphics[scale=0.55]{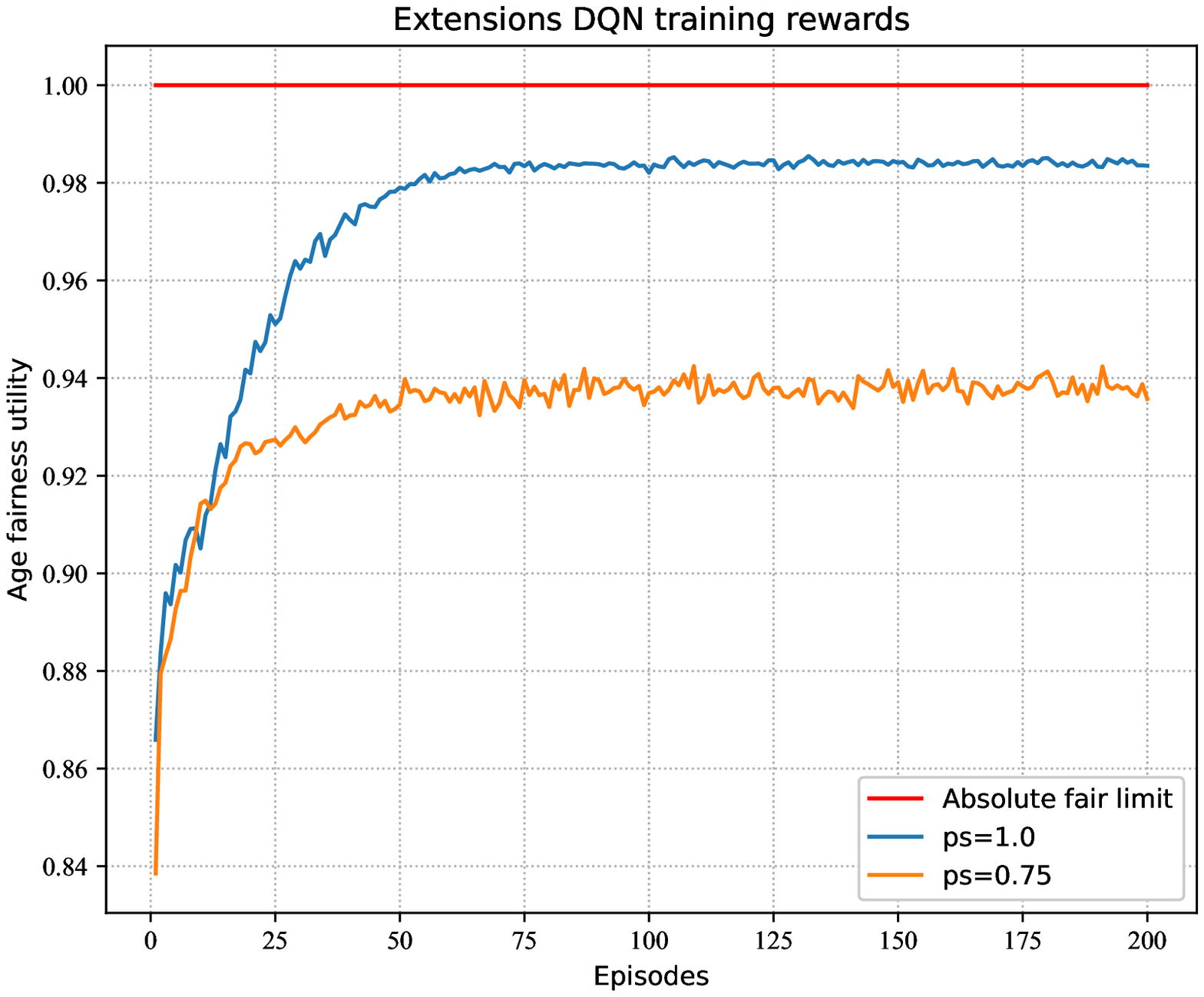} }


\centerline{\footnotesize\begin{tabular}{c} Fig.\ 4.\ Learning cuve of simple vehicle scenario
\end{tabular}}

\vskip 0.5\baselineskip
In Fig.4, the red straight line represents the absolute fairness limit, because according to Eq.(4), when the age is completely fair, the fairness loss should be 0.\ Therefore, according to Eq.(5), for each step (i.e., one time interval) of each episode, the age fairness utility is always 1.\ However, in the real environment, even if different vehicles use the same MCW, their ages cannot be completely equal because of the randomness of the integer selection of the back-off counter in the exponential back-off mechanism.\ Therefore, we only use it to express the absolute fairness upper limit.\ In the simulation analysis of the following training stage and testing stage, we will no longer describe the absolute fairness limit.

Fig.5 shows the learning curve of the training process in the complex vehicle scenario.\ It can be seen that the tendency of the curve is similar to that in the simple vehicle scenario, which means that the proposed scheme is suitable for different dynamic scenarios, except that the average age fairness utility surged from episode 0 to 100.\ 
\centerline{\includegraphics[scale=0.55]{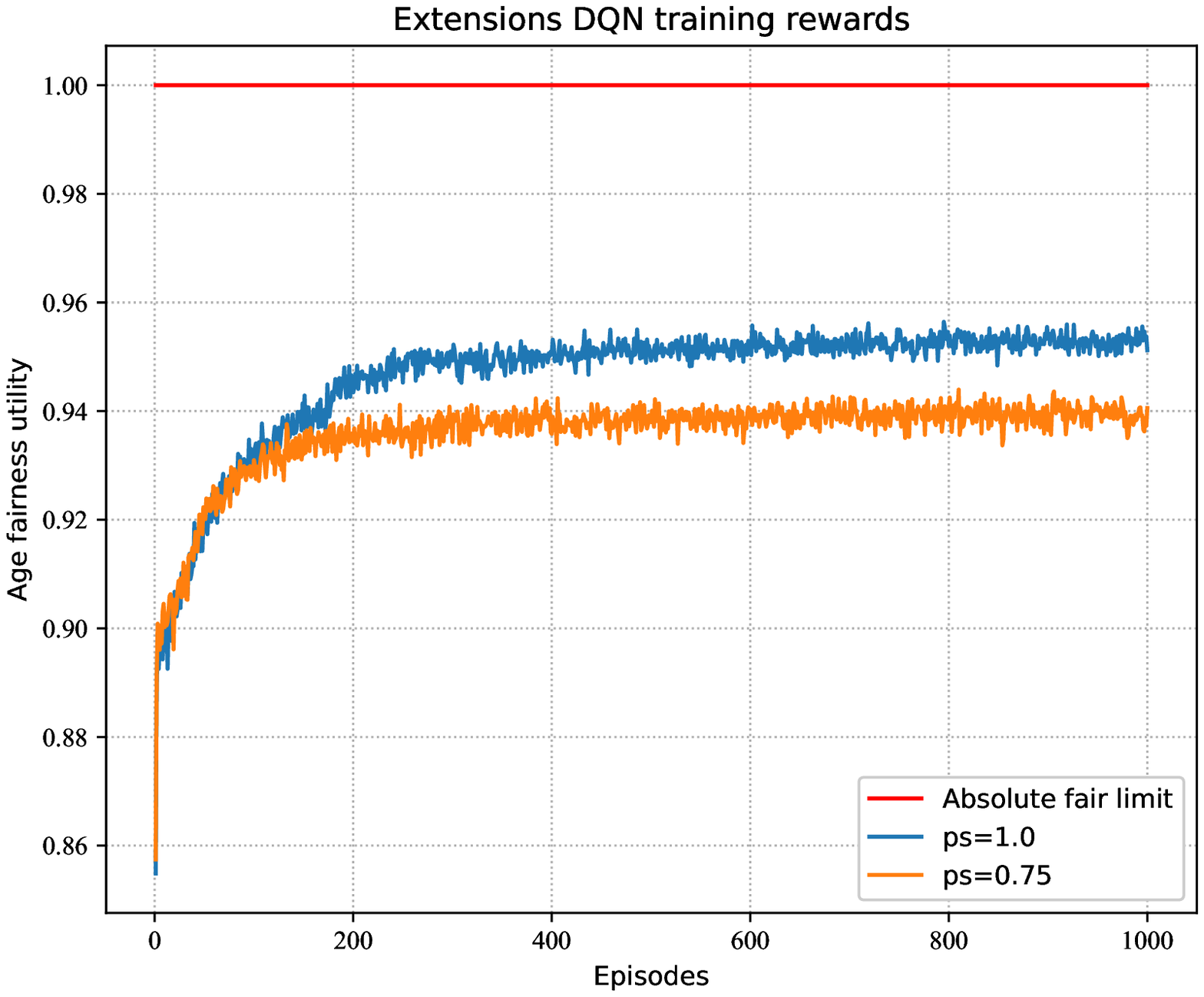} }


\centerline{\footnotesize\begin{tabular}{c} Fig.\ 5.\ Learning cuve of complex vehicle scenario
\end{tabular}}

\vskip 0.5\baselineskip
\noindent The curve increases slowly from episode 100 to 500 and reaches relative stability from 500 to 1000.\ The difference is that when $ps=1.0$, the age fairness utility after convergence is lower than that in the simple vehicle scenario.\ This is mainly because the MCWs of all vehicles in complex vehicle scenario will be more difficult to predict than that in simple vehicle scenario.\ However, when $ps=0.75$, the age fairness utility decreases slightly compared with $ps=1.0$, which means that the proposed scheme has similar adaptability to some complex scenarios.\ In addition, compared with the simple vehicle scenario, the age fairness utility of complex vehicle scenario converges at around 500 episodes, which also indicates that more model training time is required for the complex vehicle scenario.

{\bf 2.\ Testing stage}

In the testing stage, $node \ 0$ adopts the trained model to achieve the age fairness.

Fig.6(a) and (b) show the test results of age fairness utility in simple vehicle scenarios.\ To restore the real scenario, we only executed the simulation in different situations for one time, without averaging the age fairness utility.\ Fig.6(a) is the test result of 200 episodes (each episode contains 200 steps, \emph{i.e.}, 200 time intervals).\ It can be seen that during the test, for both $ps=1.0$ or $ps=0.75$, the age fairness utility converges similar to that in Fig.4, which proves the reliability of our model.\ In addition, Fig.6(b) shows the test result of the age fairness utility in an episode for the simple vehicle scenario with $ps=1.0$.\ The yellow line is the average age fairness utility of 200 steps.\ Since we are using the training dataset of the real protocol simulation, we can see that the utility of most steps fluctuate around the average value.\ It is worth noting that the age fairness utility of some steps is 1, which means that only $node \ 0$ exists in the network at this time; thus there is no fairness problem or we can also say that it is absolutely fair for the network.\ For individual points where the age fairness utility is less than 0.95, such as 90, 119, 175, 199, $node \ 0$ may make an error in predicting the MCW, which caused a sudden decrease in the age fairness utility, but we can see that we still get a relatively high average age fairness utility.
\centerline{\includegraphics[scale=0.65]{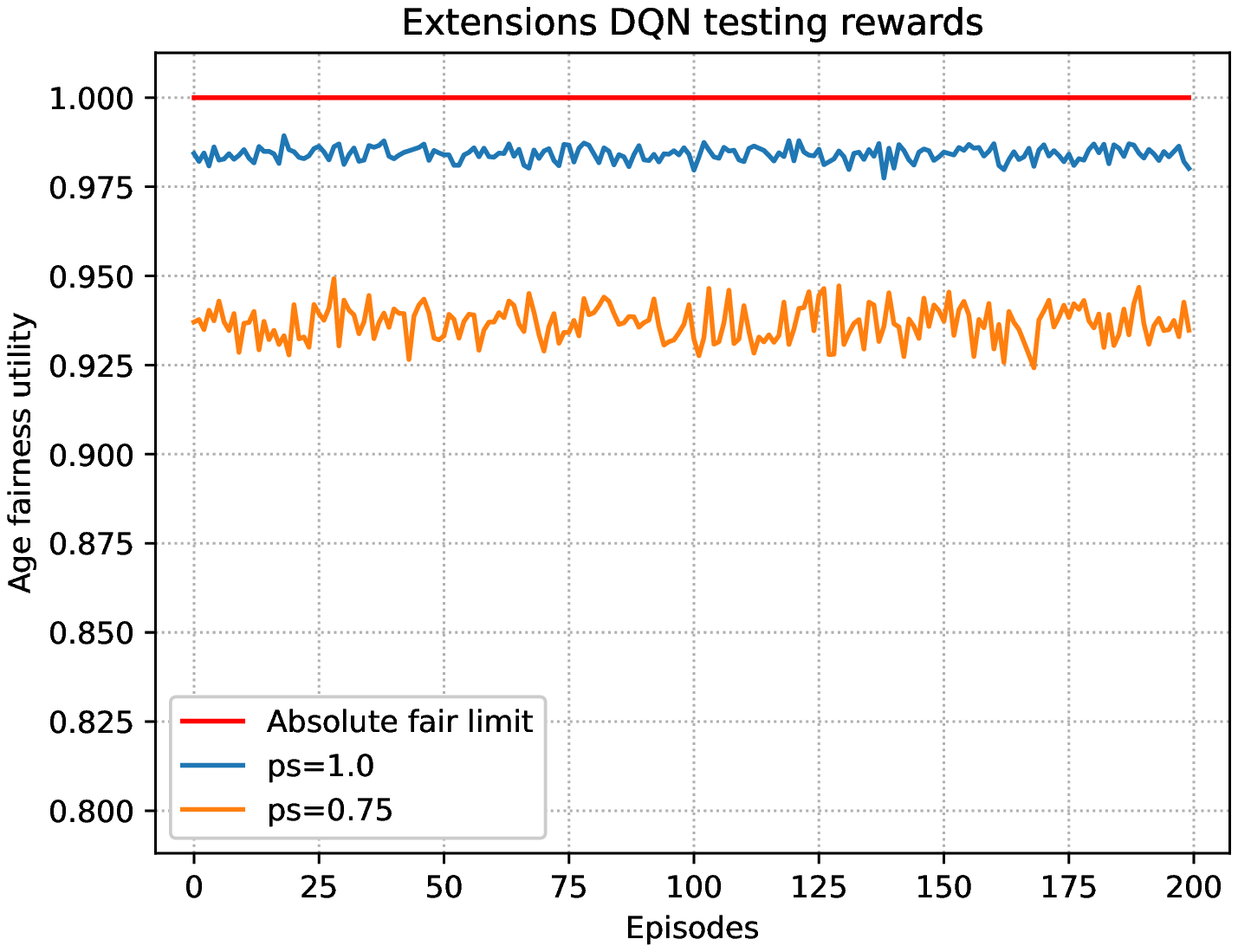} }
\centerline{\footnotesize\begin{tabular}{c} (a)
\end{tabular}}
\centerline{\includegraphics[scale=0.65]{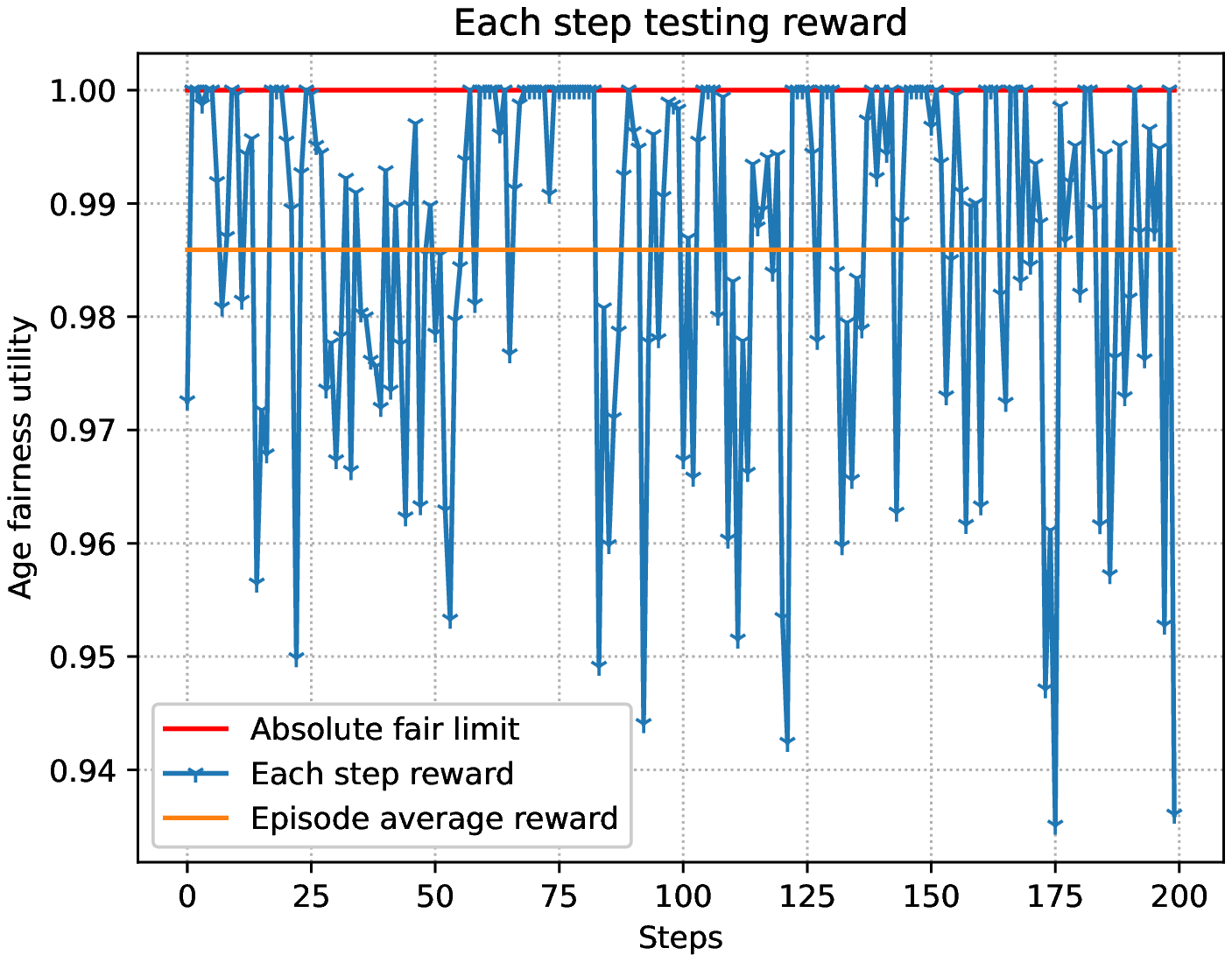} }
\centerline{\footnotesize\begin{tabular}{c} (b)
\end{tabular}}
\centerline{\footnotesize\begin{tabular}{c} Fig.\ 6.\ Testing cuve of simple vehicle scenario
\end{tabular}}
\vskip 0.5\baselineskip
Fig.7(a) and (b) show the test results of age fairness utility in complex vehicle scenarios.\ Fig.7(a) is the test result of 1000 episodes (each episode contains 200 steps, \emph{i.e.}, 200 time intervals). It can be seen that during the test, for both $ps=1.0$ or $ps=0.75$, the age fairness utility converges similar to that in Fig.5, which proves the reliability of our model.\ Different from Fig.6(a), the age fairness utility in Fig.7(a) fluctuates more greatly, and when $ps=1.0$, the age fairness utility is lower than that in Fig.6(a).\ This is mainly because for scenarios where MCW changes are more complex, the accuracy of MCW prediction will be degraded, but this does not affect the superiority of our method.\ Fig.7(b) shows the test result of the age fairness utility of 200 steps in an episode of the complex vehicle scenario with $ps=1.0$.\ In addition, our description of the curve in Fig.7(b) is the same as that in Fig.6(b).\ It can be seen that Fig.7(b) has a lower average age fairness utility than Fig.6(b) (\emph{i.e.}, yellow straight line in Fig.6(b)).\ It can be seen that it has gone through more steps before the age fairness utility decreases significantly.\ It is mainly because the MCW changes for the more complex vehicle scenario are more complicated, which will degrade the accuracy of $node \ 0$ predictions of the MCW, thus reducing the age fairness utility.\ However, it can be seen that the age fairness utility in the steps is still around the average value.\ It indicates that although the performance has been slightly reduced, $node \ 0$ can still adapt to the complex scenario.
\centerline{\includegraphics[scale=0.65]{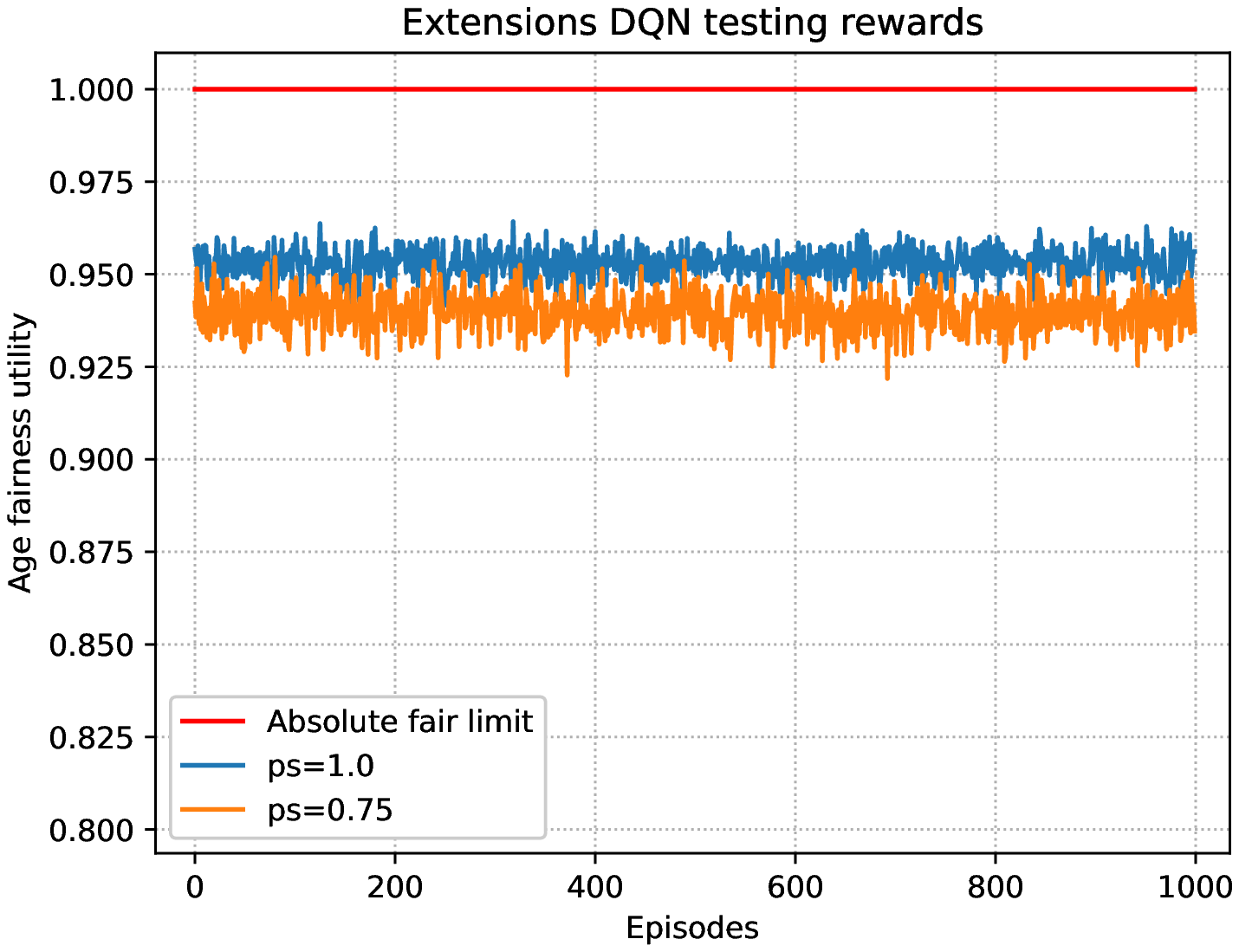} }
\centerline{\footnotesize\begin{tabular}{c} (a)
\end{tabular}}
\centerline{\includegraphics[scale=0.65]{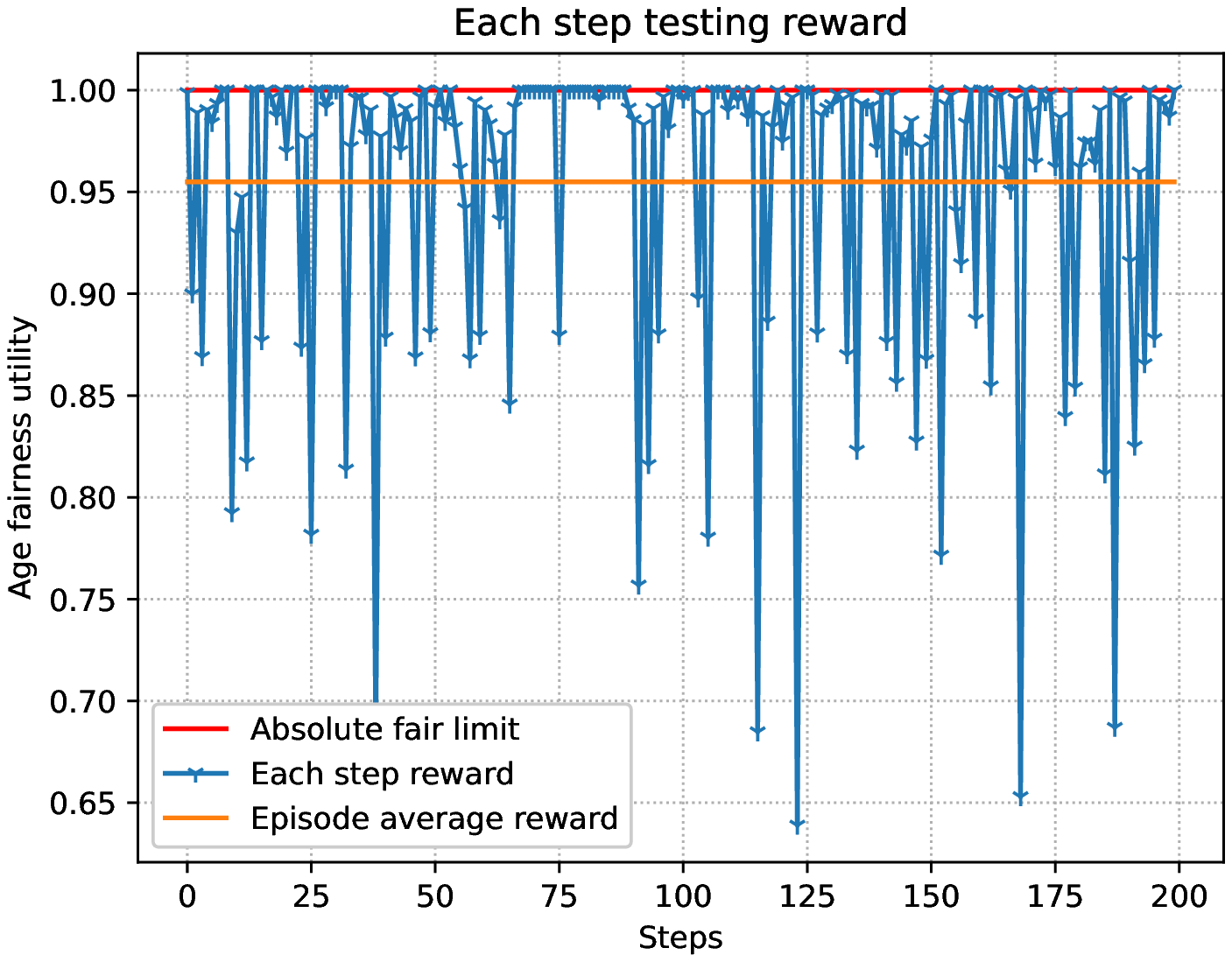} }
\centerline{\footnotesize\begin{tabular}{c} (b)
\end{tabular}}
\centerline{\footnotesize\begin{tabular}{c} Fig.\ 7.\ Testing cuve of complex vehicle scenario
\end{tabular}}
\vskip 0.5\baselineskip

{\bf 3.\ Vehicle characteristic analysis}

We simulate a complex vehicle scenario and average the test results of 1000 episodes.\ Fig.8 shows the age fairness utility when the maximum number of vehicles changes.\ The vehicle arrival rate and departure rate are equal to 3.\ When the number of vehicles in the network is greater than 1, \emph{i.e.}, there are vehicles in the network, for both $ps=1.0$ or $ps=0.75$, we can see that different numbers of vehicles may incur similar age fairness utility.\ Ideally, they should have exactly the same age fairness utility.\ But as we described earlier, we calculate the age fairness utility based on the age dataset obtained by the actual protocol simulation, there will be slight errors in the case of different numbers of vehicles, so the complete equality cannot be achieved.

\centerline{\includegraphics[scale=0.65]{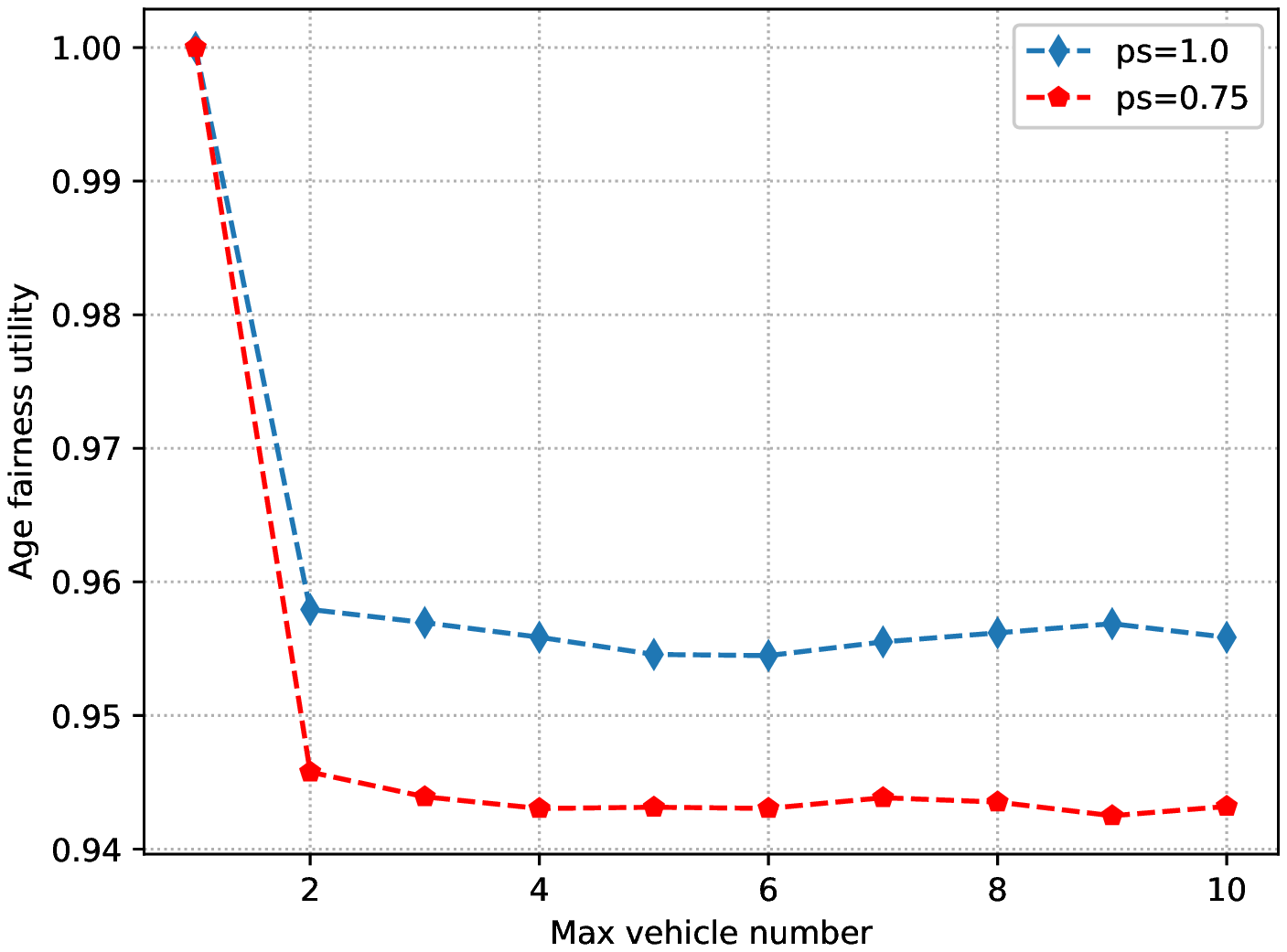} }

\centerline{\footnotesize\begin{tabular}{c} Fig.\ 8.\ Different max vehicle number rewards cuve of complex vehicle scenario
\end{tabular}}

\vskip 0.5\baselineskip
Fig.9 shows the age fairness utility under different arrival rates when the maximum number of vehicles is 6 and the departure rate is 3.\ We can see that in each observation interval $n$, due to constant departure rate, the number of vehicles in the network is very small when the arrival rate is very low.\ As there is only $node \ 0$, it is absolutely fair for the network.\ However, as the arrival rate increases, the number of vehicles in each time interval $n$ in the network gradually increases, thus leading to degradation of the age fairness utility.\ But the curve will eventually stabilize.\ The jitter ranging from 3 to 6 is attributed to the fact that age data is derived from the real protocol simulation.
\centerline{\includegraphics[scale=0.65]{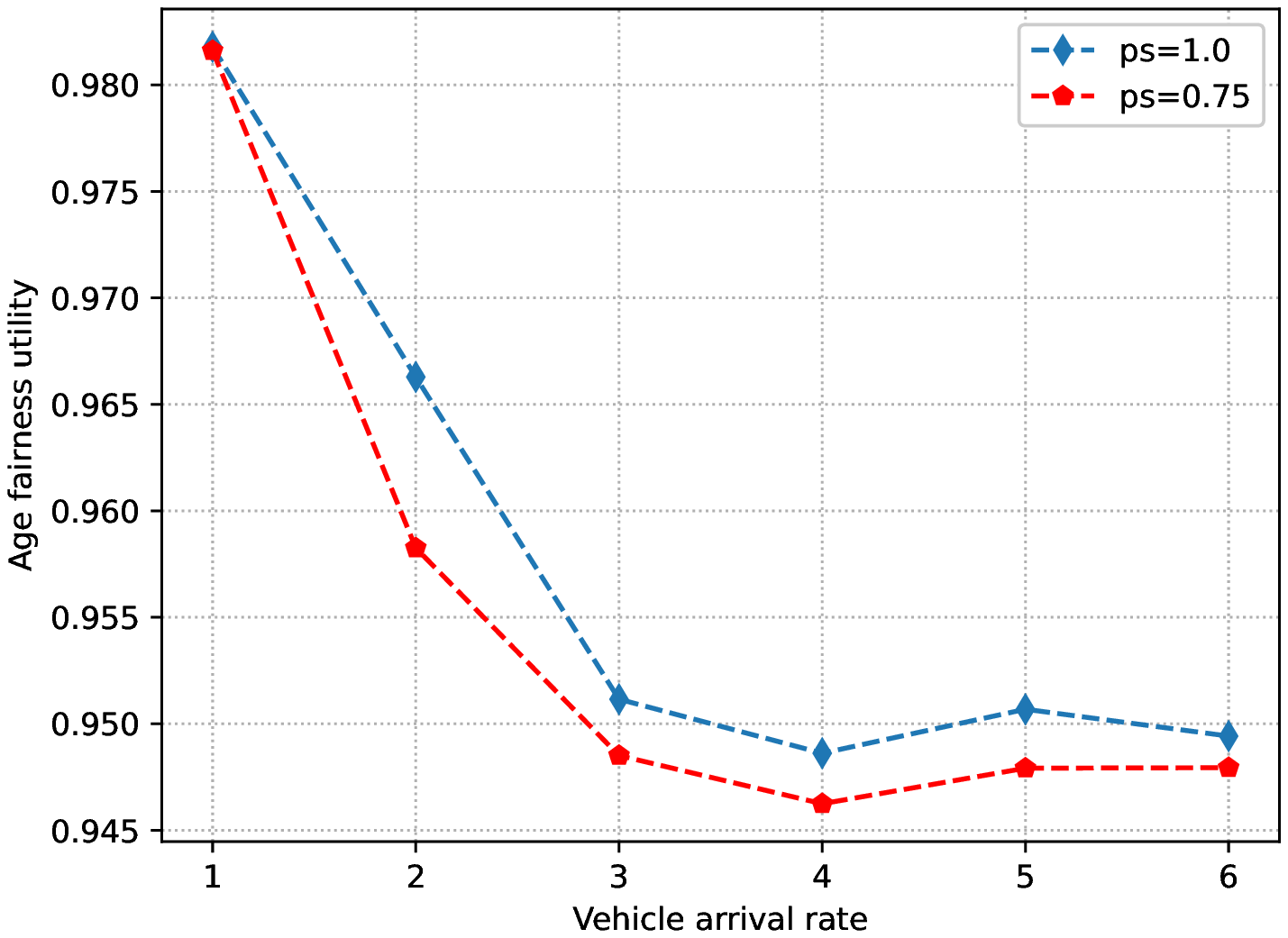} }

\centerline{\footnotesize\begin{tabular}{c} Fig.\ 9.\ Different arrival rate rewards cuve of complex vehicle scenario
\end{tabular}}

\vskip 0.5\baselineskip
Fig.10 shows the age fairness utility under different departure rates when the maximum number of vehicles is 6 and the arrival rate is 3.\ We can see that in each observation interval $n$, due to the constant arrival rate, there are many vehicles in the network when the departure rate is very low.\ However, as the departure rate increases, the number of vehicles in each time interval $n$ in the network gradually decreases, thus leading to an increase in the age fairness utility.\ As the departure rate reaches 6, there will be only $node \ 0$ in the network for a long time, which is absolutely fair for the network, and thus the age fairness utility reaches the maximum.
\centerline{\includegraphics[scale=0.65]{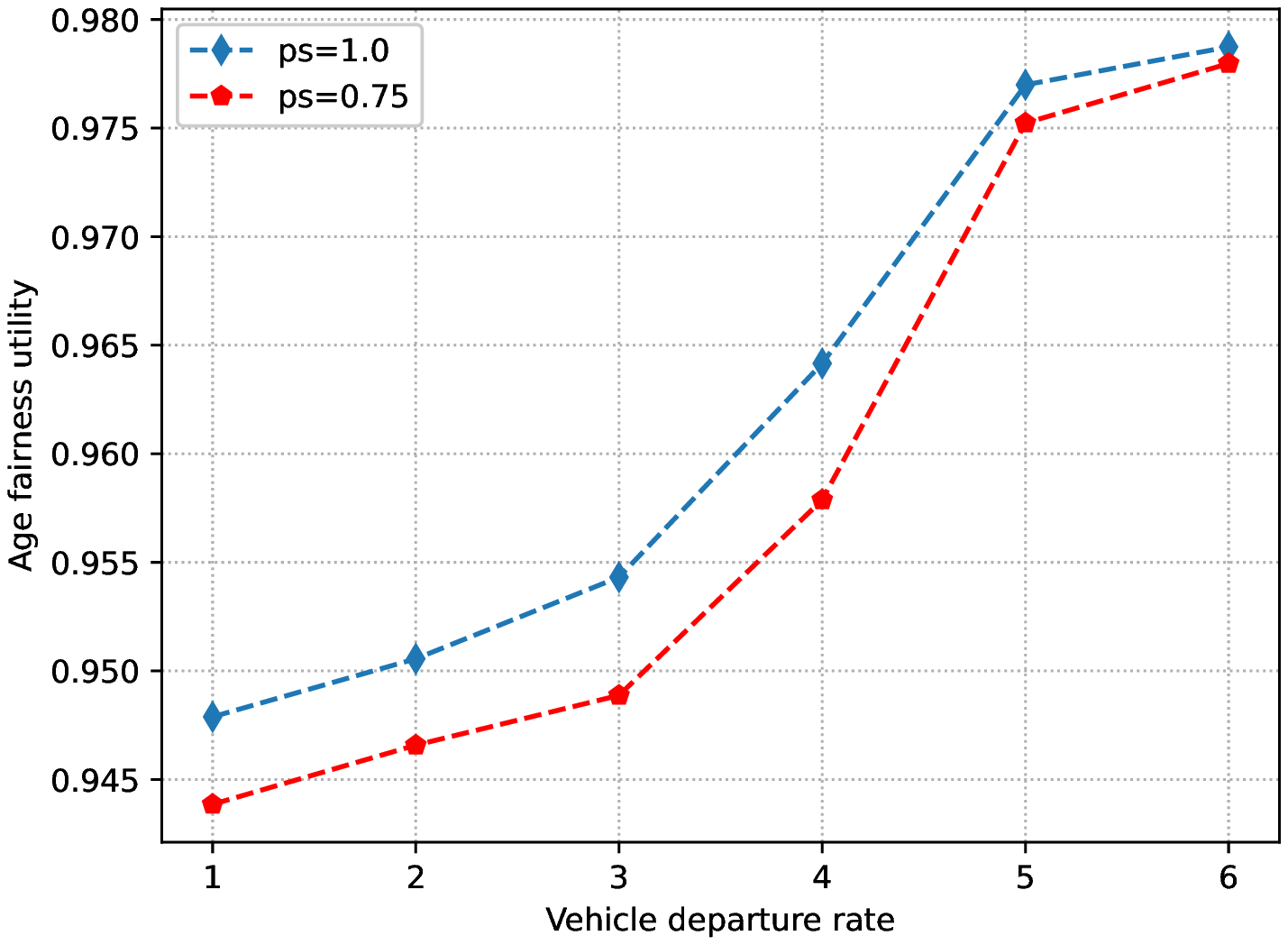} }

\centerline{\footnotesize\begin{tabular}{c} Fig.\ 10.\ Different departure rate rewards cuve of complex vehicle scenario
\end{tabular}}

\vskip 0.5\baselineskip

{\bf 4.\ Performance evaluation}

We set the maximum number of vehicles to 6, the arrival rate, and the departure rate to 3, and evaluate the performance of our RL method and the following baseline methods in two scenarios.

\textbf{Optimal (OPT):} This method is for $node \ 0$ which is fully aware of changes in the network environment including the MCW changes of other vehicles.\ However, it should be noted that this is impractical in reality, so we only use this method as the upper limit of our performance evaluation.

\textbf{Random Forest (RF):} Random forest is a supervised machine learning algorithm.\ The classifier takes local observations $[\boldsymbol{\overline{\Delta}}^n,\omega_0^n]$ as input and $\omega_0^n$ as the target label for training.\ The training label $\omega_0^{(n+1)}$ is used as the optimal action for the next step.\ We fixed the number of trees to 20 and the depth of each tree to 15.

\textbf{Decision Tree (DT):} Decision tree is also a supervised machine learning algorithm.\ Random forest algorithm just uses multiple randomly generated decision trees to generate the final output result.\ This classifier also takes local observations $[\boldsymbol{\overline{\Delta}}^n,\omega_0^n]$ as input and $\omega_0^n$ as the target label for training.\ We set the depth of the tree to 20.

\textbf{Standard Protocol (SP):} The intelligent $node \ 0$ follows the fixed MCW protocol mentioned in the IEEE 802.11 DCF protocol$^{[43]}$, and its MCW is fixed to 64 and 128 respectively in simple vehicle scenarios.\ In complex vehicle scenes, MCW is fixed to 64, 128, 256, and 512, respectively.

Fig.11(a) shows the comparison of the age fairness utility between our RL method and other baseline methods in the simple vehicle scenario.\ Among them, the red and blue box-whisker plots represent the cases of $ps=1.0$ and $ps=0.75$, respectively.\ We can see that no matter what the situation is, the age fairness utility achieved by our RL method is always the highest, and the fluctuation range of the box-whisker plot is also the smallest, (except for the OPT method), which means that the average age fairness utility of each episode is very stable.\ When $ps=1.0$, the performance of RF is slightly worse than that of RL, but it also achieves a relatively ideal age fairness utility.\ But as compared to RF, the DT's performance will be slightly reduced, when $ps=0.75$; however, RL can still learn a better policy.\ In addition, as compared to OPT, RL only slightly extends the fluctuation range, while RF and DT not only decrease the performance in terms of average age fairness utility, but also extend the fluctuation range of the box-whisker plot.\ Generally speaking, the age fairness utility achieved by the RL method is slightly better than that of DT method.\ Nevertheless, the above three methods are better than the SP method.\ For the SP method, we can see that the change of $ps$ has little impact on the age fairness utility, because in either case, $node \ 0$ uses a fixed MCW and the box-whisker plots of $ps=1.0$ and $ps=0.75$ are almost equal.\ For MCW = 64, the age fairness utility is higher than that when MCW = 128.\ This is because for simple vehicle scenario, the MCW state space of all vehicles is $[32,128]$.\ When the MCW of other vehicles is 32, the opportunity for $node \ 0$ to access the channel will be reduced, and the age fairness utility of $node \ 0$ will be reduced.\ When the MCW of all vehicles becomes 128, the opportunity of $node \ 0$ to access the channel will increase, and thus the age fairness utility of $node \ 0$ will increase while the age fairness utility of the whole simulation process will not be too low.\ When the MCW of $node \ 0$ is 128, $node \ 0$ will have less opportunity to access the channel for a long time, thus incurring a lower age fairness utility of the whole simulation process than that when MCW = 64.

Fig.11(b) shows the comparison of the age fairness utility between our RL method and other baseline methods in the complex vehicle scenario.\ In complex vehicle scenario, the MCW of all vehicles has a larger state space $[32,64,128,256,512]$, and the changes are more complex, as described at the beginning of this section. 

Fig.11(b) shows that RL achieves the result closest to the optimal utility among all the methods considering the two values of $ps$.\ As expected, the case of $p = 0.75$ is more challenging.\ Also note that the performance gap between OPT and RL is reasonable.\ Because the former has complete knowledge of the environment, while the latter must rely entirely on local observations to gather knowledge of the environment.\ In addition, the performance of RL and DT methods has been greatly reduced.\ When $ps=1.0$, it can only be the same as the performance when the MCW is 64 and 128, and when $ps=0.75$, its performance is significantly degraded.\ This is because for more complex situations, it is difficult for the classifier to give the best action to the current state of the network.\ Sometimes, the action given by the classifier may be the worst action, which will cause the average age fairness utility to be significantly degraded or even worse than the standard protocol.\ For the SP method, when the MCW used by $node \ 0$ increases, from the perspective of $node \ 0$, the age fairness utility will gradually decrease.\ This is because the opportunities for $node \ 0$ to access the channel will decrease, leading to a sharp increase in terms of the age of $node \ 0$. So far it also further reflects the effectiveness of the proposed method.
\centerline{\includegraphics[scale=0.65]{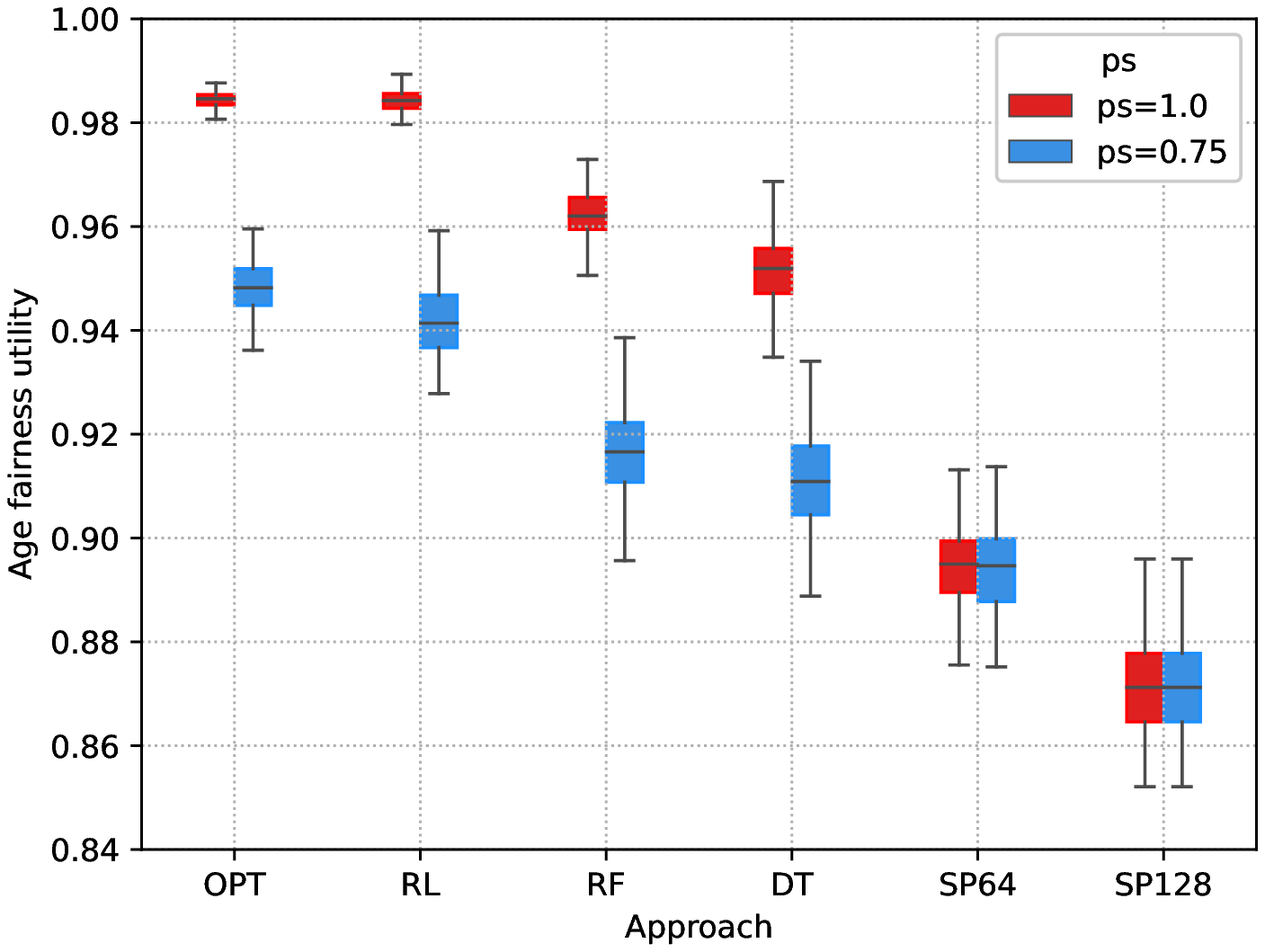} }

\centerline{\footnotesize\begin{tabular}{c} (a)
\end{tabular}}
\centerline{\includegraphics[scale=0.65]{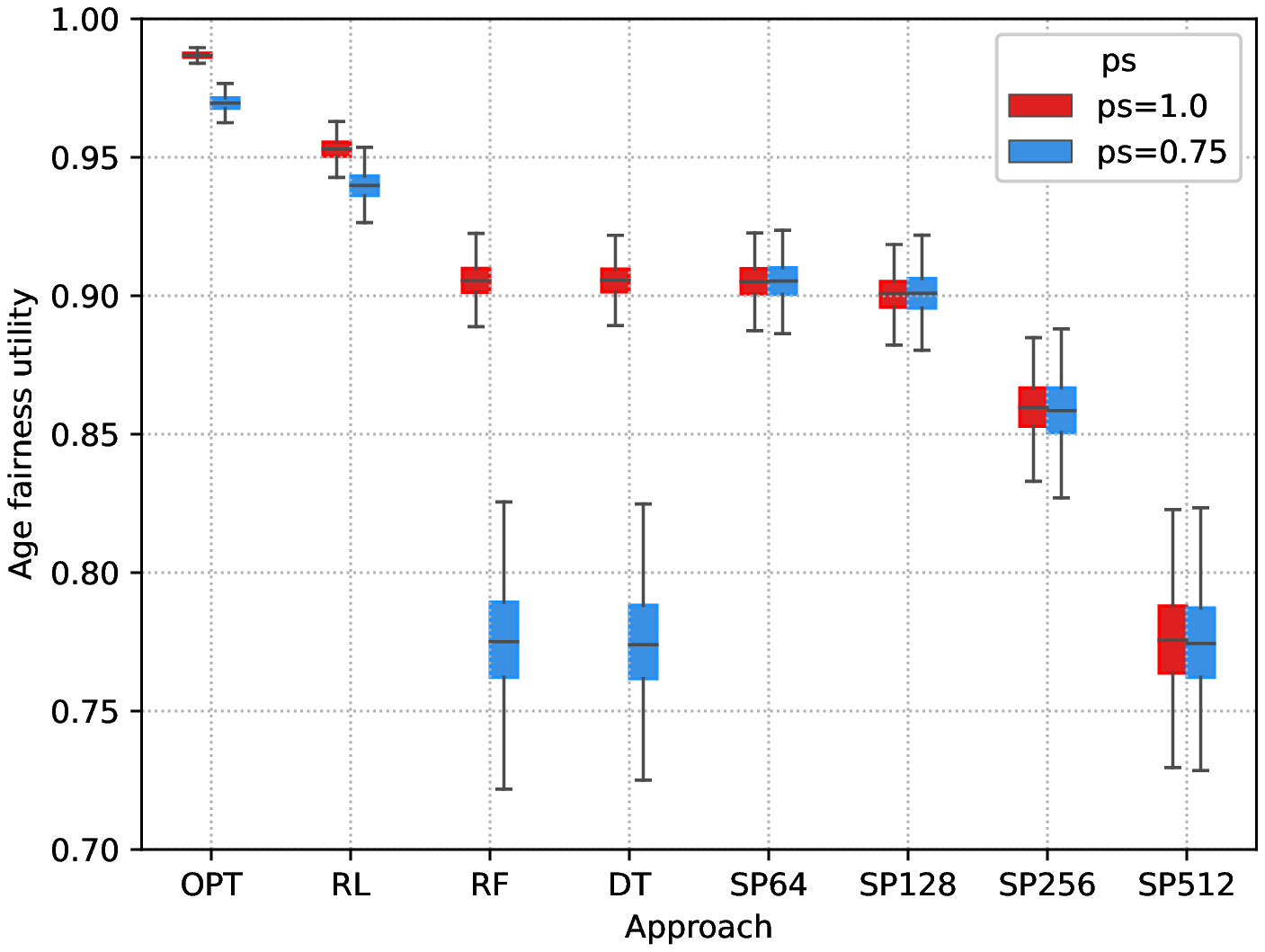} }

\centerline{\footnotesize\begin{tabular}{c} (b)
\end{tabular}}
\centerline{\footnotesize\begin{tabular}{c} Fig.\ 11.\ Performance evaluation of the proposed RL method (a)simple vehicle scenario; (b)complex vehicle scenario
\end{tabular}}

\vskip 0.5\baselineskip

\

\begin{center}{\large\bf VII.\ Conclusion and Future Work
}\end{center}

In this paper, we considered a dynamic and uncertain V2I communication scenario, where each vehicle may change its MCW to achieve more access opportunities at the expense of others and each vehicle is not willing to share its MCW with other vehicles. In this scenario, we designed an intelligent vehicular node that is used to learn the dynamics and predict the optimal MCW from local observations to ensure its age fairness. In order to allocate the optimal MCW for the vehicular node, we proposed a learning algorithm by extending the traditional DQN training and testing method to make a desirable decision by learning from replay history data.\ In addition, we obtain an age dataset for model training through real-time protocol simulation, and the superiority of our proposed RL method is proved through experiment simulations.\ According to theoretical analysis and simulation experiment, we can get the following conclusions:

{\footnotesize{$\bullet$}}\ For different state transition probabilities $ps$, the model has excellent adaptability and can achieve relatively high age fairness utility.

{\footnotesize{$\bullet$}}\ For the maximum number of vehicles in different networks, the model can also achieve approximately the same age fairness utility.

{\footnotesize{$\bullet$}}\ In order to maximize long-term discount rewards, given different vehicle arrival rates and departure rates, this method can also achieve approximately the same age fairness utility.

If each vehicle is allowed to become an intelligent vehicle, it has the ability to independently learn and adjust its own MCW, so that it can realize the adaptive compatibility problem of various network environments.\ This may lead to huge training difficulties.\ In the future, we will further study the multi-agent training problem for the adaptive MCW in the network.

\RE

\footnotesize\rm

\REF{[1]} Q. Wu, H. X. Liu, C. Zhang, \emph{et al}., ``Trajectory Protection Schemes Based on a Gravity Mobility Model in IoT,'' {\it Electronics}, vol.8, no.148, 2019.

\REF{[2]} S. Wan, J. Lu, P. Fan, \emph{et al}., ``To Smart City: Public Safety Network Design for Emergency,'' {\it IEEE Access}, vol.6, pp.1451-1460, 2017.

\REF{[3]} H. B. Zhu, Q. Wu, X. Wu, \emph{et al}., ``Decentralized Power Allocation for MIMO-NOMA Vehicular Edge Computing Based on Deep Reinforcement Learning,'' {\it IEEE Internet of Things Journal}, DOI: 10.1109/JIOT.2021.3138434, 2021.

\REF{[4]} J. H. Zhao, Q. P. Li, Y. Gong, \emph{et al}., ``Computation offloading and resource allocation for cloud assisted mobile edge computing in vehicular networks,'' {\it IEEE Transactions on Vehicular Technology}, vol.68, no.8, pp.7944-7956, 2019.

\REF{[5]} Q. Wu and J. Zheng, ``Performance Modeling and Analysis of the ADHOC MAC Protocol for VANETs,'' in {\it Proc. of  IEEE International Conference on Communication (ICC)}, London, UK, pp.3646-3652, 2015.

\REF{[6]} Q.\ Wu, Y.\ Zhao and Q.\ Fan, ``Time-Dependent Performance Modeling for Platooning Communications at Intersection,'' {\it IEEE Internet of Things Journal}, 2022.

\REF{[7]} Defense Advanced Research Projects Agency, ``Spectrum Collaboration Challenge (SC2),''.

\REF{[8]} J. H. Zhao, L. H. Yang, M. H. Xia, \emph{et al}., ``Unified Analysis of Coordinated Multi-Point Transmissions in mmWave Cellular Networks,'' {\it IEEE Internet of Things Journal}, DOI: 10.1109/JIOT.2021.3134017, 2021.

\REF{[9]} Q. Wu and J. Zheng, ``Performance Modeling and Analysis of the ADHOC MAC Protocol for Vehicular Networks,'' {\it Wireless Networks}, vol.22, no.3, pp.799-812, 2016.

\REF{[10]} J. H. Zhao, X. K. Sun, Q. P. Li, \emph{et al}., ``Edge Caching and Computation Management for Real-Time Internet of Vehicles: An Online and Distributed Approach,'' {\it IEEE Transactions on Intelligent Transportation Systems}, vol.22, no.4, pp.2183-2197, 2021.

\REF{[11]} Z. Yao, J. Jiang, P. Fan, \emph{et al}., ``A Neighbor-table-based Multipath Routing in Adhoc Networks,'' in {\it Proc. of the 57th IEEE Semiannual Vehicular Technology Conference}, VTC 2003-Spring.. Vol. 3. IEEE, 2003.

\REF{[12]} Q.\ Wu, H. M. Ge, P. Y. Fan, \emph{et al}., ``Time-dependent Performance Analysis of the 802.11p-based Platooning Communications Under Disturbance,'' {\it IEEE Transactions on Vehicular Technology}, vol.69, no.12, pp.15760-15773, 2020.

\REF{[13]} Q. Wang, D O. Wu, P. Fan. ``Delay-constrained Optimal Link Scheduling in Wireless Sensor Networks,'' {\it IEEE Transactions on Vehicular Technology}, vol.59, no.9, pp.4564-4577, 2010.

\REF{[14]} X. Chen, J. Lu, P. Fan, \emph{et al}., ``Massive MIMO Beamforming with Transmit Diversity for High Mobility Wireless Communications,'' {\it IEEE Access}, vol.5, pp.23032-23045, 2017.

\REF{[15]} K. Wang, FR. Yu, L. Wang, \emph{et al}., ``Interference Alignment with Adaptive Power Allocation in Full-duplex-enabled Small Cell Networks,'' {\it IEEE Transactions on Vehicular Technology}, vol.68, no.3, pp.3010-3015, 2019.

\REF{[16]} J. Fan, Q. Wu, JF. Hao. ``Optimal Deployment of Wireless Mesh Sensor Networks Based on Delaunay Triangulations  in {\it Proc. of International Conference on Information}, Networking and Automation (ICINA). IEEE, 2010, 1: V1-370-V1-374.

\REF{[17]} R.\ D.\ Yates, Y.\ Sun, D.\ R.\ Brown, \emph{et al}., ``Age of Information: An Introduction and Survey,'' {\it IEEE Journal on Selected Areas in Communications}, vol.39, no.5, pp.1183-1210, 2021.

\REF{[18]} X. F. Chen, C. Wu, T. Chen, \emph{et al}., ``Age of Information Aware Radio Resource Management in Vehicular Networks: A Proactive Deep Reinforcement Learning Perspective,'' {\it IEEE Transactions on Wireless Communications}, vol.19, no.4, pp.2268-2281, 2020.

\REF{[19]} M. K. Abdel-Aziz, S. Samarakoon, C. Liu, \emph{et al}., ``Optimized Age of Information Tail for Ultra-Reliable Low-Latency Communications in Vehicular Networks,'' {\it IEEE Transactions on Communications}, vol.68, no.3, pp.1911-1924, 2020.

\REF{[20]} Q.\ Wu, Z. Y. Wan, Q.\ Fan, \emph{et al}., ``Velocity-adaptive Access Scheme for MEC-assisted Platooning Networks: Access Fairness Via Data Freshness,'' {\it IEEE Internet of Things Journal}, vol.9, no.6, pp.4229-4244, 2021.

\REF{[21]} J.\ Lv, X.\ M. Zhang, X.\ J. Han, \emph{et al}., ``A Novel Adaptively Dynamic Tuning of the Contention Window (CW) for Distributed Coordination Function in IEEE 802.11 Ad hoc Networks,'' in {\it Proc. of 2007 International Conference on Convergence Information Technology (ICCIT 2007)}, Gwangju, Korea (South), pp.290-294, 2007.

\REF{[22]} C.\ Wu, S.\ Ohzahata, Y.\ S. Ji, \emph{et al}., ``A MAC protocol for delay-sensitive VANET applications with self-learning contention scheme,'' in {\it Proc. of 2014 IEEE 11th Consumer Communications and Networking Conference (CCNC)}, Las Vegas, NV, USA, pp.438-443, 2014.

\REF{[23]} A.\ Jamali, S.\ M. S. Hemami, M.\ Berenjkoub, \emph{et al}., ``An adaptive MAC protocol for wireless LANs,'' {\it Journal of Communications and Networks}, vol.16, no.3, pp.311-321, 2014.

\REF{[24]} X.\ Zhou, C.\ W. Zheng and X.\ X. He, ``Adaptive contention window tuning for IEEE 802.11,'' in {\it Proc. of 2015 22nd International Conference on Telecommunications (ICT)}, Sydney, NSW, Australia, pp.74-79, 2015.

\REF{[25]} A.\ Pressas, Z.\ G. Sheng, F.\ Ali, \emph{et al}., ``Contention-based learning MAC protocol for broadcast vehicle-to-vehicle communication,'' in {\it Proc. of 2017 IEEE Vehicular Networking Conference (VNC)}, Turin, Italy, pp.263-270, 2017.

\REF{[26]} C.\ Wu, Z.\ Liu, F.\ Q. Liu, \emph{et al}., ``Collaborative Learning of Communication Routes in Edge-Enabled Multi-Access Vehicular Environment,'' {\it IEEE Transactions on Cognitive Communications and Networking}, vol.6, no.4, pp.1155-1165, 2020.

\REF{[27]} X.\ W. Wu, X.\ H. Li, J.\ Li, \emph{et al}., ``Deep Reinforcement Learning for IoT Networks: Age of Information and Energy Cost Tradeoff,'' in {\it Proc. of GLOBECOM 2020 - 2020 IEEE Global Communications Conference}, Taipei, Taiwan, pp.1-6, 2020.

\REF{[28]} X.\ F. Chen, C.\ Wu, T.\ Chen, \emph{et al}., ``Age of Information Aware Radio Resource Management in Vehicular Networks: A Proactive Deep Reinforcement Learning Perspective,'' {\it IEEE Transactions on Wireless Communications}, vol.19, no.4, pp.2268-2281, 2020.

\REF{[29]} F.\ Y. Wu, H.\ L. Zhang, J.\ J. Wu, \emph{et al}., ``UAV-to-Device Underlay Communications: Age of Information Minimization by Multi-Agent Deep Reinforcement Learning,'' {\it IEEE Transactions on Communications}, vol.69, no.7, pp.4461-4475, 2021.

\REF{[30]} S.\ H. Wang, M.\ Z. Chen, W. Saad, \emph{et al}., ``Reinforcement Learning for Minimizing Age of Information in Real-time Internet of Things Systems with Realistic Physical Dynamics,'' in {\it Proc. of GLOBECOM 2020 - 2020 IEEE Global Communications Conference}, Taipei, Taiwan, pp.1-6, 2020.

\REF{[31]} B.\ Han, Y.\ Zhu, Z.\ Y. Jiang, \emph{et al}., ``Fairness for Freshness: Optimal Age of Information Based OFDMA Scheduling with Minimal Knowledge,'' {\it IEEE Transactions on Wireless Communications}, vol.20, no.12, pp.7903-7919, 2021.

\REF{[32]} S.\ Y. Leng and A.\ Yener, ``An Actor-Critic Reinforcement Learning Approach to Minimum age of Information Scheduling in Energy Harvesting Networks,'' in {\it Proc. of ICASSP 2021 - 2021 IEEE International Conference}, Toronto, ON, Canada, pp.8128-8132, 2021.

\REF{[33]} R.\ D.\ Yates, Y.\ Sun, D.\ R.\ Brown, \emph{et al}., ``Age of Information: An Introduction and Survey,'' {\it IEEE Journal on Selected Areas in Communications}, vol.39, no.5, pp.1183-1210, 2021.

\REF{[34]} I.\ Kadota and E.\ Modiano, ``Minimizing the Age of Information in Wireless Networks with Stochastic Arrivals,'' {\it IEEE Transactions on Mobile Computing}, vol.20, no.3, pp.1173-1185, 2021.

\REF{[35]} G.\ Bianchi, ``Performance analysis of the IEEE 802.11 distributed coordination function,'' {\it IEEE Journal on Selected Areas in Communications}, vol.18, no.3, pp.535-547, 2000.

\REF{[36]} R.\ S. Sutton and A.\ G. Barto, ``Introduction to Reinforcement Learning,'' {\it MIT press Cambridge}, vol.135, 1998.

\REF{[37]} V.\ Hasselt, A.\ Guez and D.\ Silver, ``Deep Reinforcement Learning with Double Q-learning,'' in {\it Proc.\ of the AAAI Conference on Artificial Intelligence}, vol.30, no.1, 2016.

\REF{[38]} Y.\ Zhang, P.\ Sun, Y.\ H. Yin, \emph{et al}., ``Human-like Autonomous Vehicle Speed Control by Deep Reinforcement Learning with Double Q-Learning,'' in {\it Proc. of 2018 IEEE Intelligent Vehicles Symposium (IV)}, Changshu, China, pp.1251-1256, 2018.

\REF{[39]} Z.\ Y. Wang, T.\ Schaul, M.\ Hessel, \emph{et al}., ``Dueling Network Architectures for Deep Reinforcement Learning'' in {\it Proc.\ of The 33rd International Conference on Machine Learning}, vol.48, pp.1995-2003, 2016.

\REF{[40]} M.\ Hessel, J.\ Modayil, H.\ V.\ Hasselt, \emph{et al}., ``Rainbow: Combining Improvements in Deep Reinforcement Learning,'' in {\it Proc. of Thirty-second AAAI conference on artificial intelligence}, 2018.

\REF{[41]} M. Fortunato, M. G. Azar, B. Piot, \emph{et al}., ``Noise Networks for Exploration.'' {\it arXiv preprint}, arXiv:\ 1706.10295, 2017.

\REF{[42]} H. V.\ Hasselt, A.\ Guez, M.\ Hessel, \emph{et al}., ``Learning Functions Across Many Orders of Magnitudes,'' in {\it Proc.\ of the Advances in Neural Information Processing Systems}, Barcelona, Spain, pp.80-99, 2016.

\REF{[43]} Q. Wu, S. Xia, P. Fan, \emph{et al}., ``Velocity-adaptive V2I Fair-access Scheme Based on IEEE 802.11 DCF for Platooning Vehicles,'' {\it Sensors}, vol.18, no.12, pp.4198, 2018.

\end{document}